\documentclass[12pt,twoside]{article}\usepackage{a4}\usepackage{amssymb}
\usepackage{graphicx}\usepackage{amsmath}
\numberwithin{equation}{section}

\newcommand{\be}{\begin{equation}}\newcommand{\ee}{\end{equation}}
\newcommand{\bea}{\begin{eqnarray}}\newcommand{\eea}{\end{eqnarray}}
\newcommand{\nn}{\nonumber}
\newcommand{\pa}{\partial}

\newcommand{\ga}{\gamma}
\newcommand{\om}{\omega}
\newcommand{\ep}{\varepsilon}
\renewcommand{\phi}{\varphi}
\newcommand{\Ref}[1]{(\ref{#1})}
\newcommand{\F}{{\cal F}}
\newcommand{\suml}{\sum_{l=0}^{\infty}{\vphantom{\sum}}^{\prime}}
\newcommand{\rE}{r_{\rm TE}}\newcommand{\rM}{r_{\rm TM}}

\renewcommand{\a}{{(a)}}\renewcommand{\b}{{(b)}}
\newcommand{\cc}{{\rm c.c.}}
\renewcommand{\r}{r_1}

\newcommand{\elm}{electromagnetic~}
\newcommand{\Om}{\omega_p}
\newcommand{\Oma}{\omega_{p\,1}}\newcommand{\Omb}{\omega_{p\,2}}
\newcommand{\Omi}{\omega_{p\,i}}

\renewcommand{\kappa}{\varkappa}
\title{Low temperature expansion in the Lifshitz formula}
\author{M. Bordag\footnote{bordag@itp.uni-leipzig.de}\\
\small Universit\"{a}t Leipzig, Institute for Theoretical Physics, Germany}
\date{\small  \today}
\begin{document}
\maketitle

\begin{abstract}
The low temperature expansion of the free energy in a Casimir effect setup is considered in detail. The starting point is the Lifshitz formula in Matsubara representation and  the basic method is its reformulation using the Abel-Plana formula making full use of the analytic properties. This provides a unified description of specific models. We re-derive the known results  and, in a number of cases, we are able to go beyond. We also discuss the cases with dissipation. It is an aim of the paper to give a coherent exposition of the asymptotic expansions for $T\to0$. The paper includes the derivations and should provide a self contained representation.
\end{abstract}
\tableofcontents
\thispagestyle{empty}
\section{Introduction}
The Lifshitz formula is the basic tool for the calculation of van der Waals and Casimir forces between two material half spaces. It emerged in 1956 for the description of the electromagnetic dispersion forces. Together with Casimir's approach of zero-point or vacuum fluctuations these are two sides of one coin. In the language of quantum field theory these are one-loop corrections to a classical background which may be given by boundary conditions or by classical fields as well.

\begin{figure}\unitlength 1cm
 \begin{picture}(8,6)
  \put(0,-3){\includegraphics[width=15 cm]{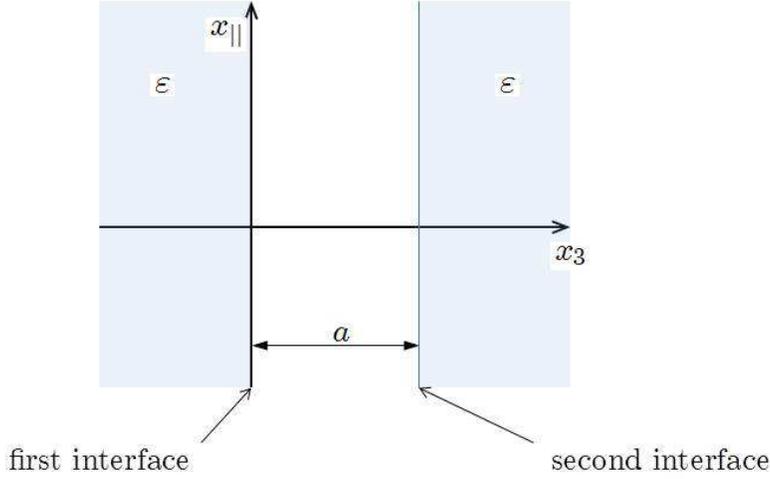}    }   \end{picture}
  \caption{The configuration of two interfaces separated by a gap of width $a$.}\label{fig1}
\end{figure}

For the configuration of two parallel interfaces with a gap of widths $a$ between them (see Fig. \ref{fig1}), the Lifshitz formula
\be\label{1.LF} \F=k_{\rm B} T\suml\int\frac{d\mathbf{k}}{(2\pi)^2}
        \sum_{\rm TE,TM}\ln\left(1-    
                                    r_1 r_2
                                            \,e^{-2a\eta}\right)
\ee
provides the separation dependent part of the free energy of the electromagnetic field at temperature $T$ in terms of the reflection coefficients $r_i$ (i=1,2) of the two interfaces. In \Ref{1.LF}, the prime on the sum indicates that the $(l=0)$-contribution must be taken with a factor $1/2$. Initially this formula was written for dielectric half spaces with permittivity $\ep$ behind the interfaces with the well known reflection coefficients
\be\label{1.r}  r_{\rm TE}=\frac{\eta-\kappa}{\eta+\kappa},    \qquad
                r_{\rm TM}=\frac{\ep\eta-\kappa}{\ep\eta+\kappa},
\ee
which must be inserted for $r_i$ according to the polarization. In \Ref{1.LF}, the integration is over the wave numbers $\mathbf{k}=\{k_1,k_2\}$ in the directions in parallel to the interfaces, the summation is over the Matsubara frequencies
\be\label{1.xi} \xi_l=2\pi k_{\rm B}T l \quad(l \mbox{ integer})
\ee
and the notations
\be\label{1.etaxi}  \eta=\sqrt{\xi_l^2/c^2+k^2},\qquad \kappa=\sqrt{\ep \xi_l^2/c^2+k^2},
\ee
($k=|\mathbf{k}|$) are used. Here, and in \Ref{1.LF}, the permittivity is allowed to be frequency dependent, $\ep=\ep(\om)$, and must be taken at imaginary frequency, $\om=i \xi_l$.

The Lifshitz formula \Ref{1.LF} is basic for the calculation of the force
$F$ acting between the interfaces (actually the pressure since \Ref{1.LF} is the free energy per unit surface),
\be\label{1.F}  F=-\frac{d}{da}\F.
\ee
At present, this force can be compared with results from force measurements on a high level of precision, see \cite{BKMM} or the recent review \cite{klim09-81-1827}. Also, the Lifshitz formula was generalized to arbitrary, non-flat geometry of the interfaces in terms of the scattering approach, see, for example, Chapter 10 in \cite{BKMM}.

Along with the great success, it must be mentioned that a decade ago a problem appeared the Lifshitz formula has when including dissipation. For instance, when inserting the permittivity of the Drude model,
\be\label{1.epDr}   \ep^{\rm Dr}(\om)=1-\frac{\Om^2}{\om(\om+i\gamma)},
\ee
with a temperature dependent dissipation parameter $\gamma(T)$, decreasing for $T\to0$ sufficiently fast (see Eq. \Ref{4.4.2.ga}), the free energy \Ref{1.F} violates the third law of thermodynamics (Nernst's heat theorem) \cite{beze04-69-022119}.
This means, the separation dependent part of the entropy,
\be\label{1.S}S=-\frac{\pa}{\pa T}\,\F,
\ee
has a non-vanishing limit for $T\to0$. The physical interpretation of this problem is still under discussion.  Another, closely related problem appears if decreasing the dissipation parameter $\gamma$ at fixed temperature. The naive expectation would be to re-obtain in the limit the free energy of the plasma model. While the permittivity \Ref{1.epDr} turns into that of the plasma model, the free energy has an additional contribution for $\gamma\to0$, i.e., it is not perturbative in $\ga$. This is counter intuitive from the physical point of view since a small dissipation should have a small effect on the dispersion force.  Also, there is growing evidence for a disagreement between the predictions following from the Lifshitz formula with dissipation and measurements \cite{klim09-81-1827,chan11-107-090403,bani12-85-045436,chan12-85-165443,bani12-85-195422}.

At finite temperature, the vacuum fluctuations appear together with the thermal fluctuations of the electromagnetic field. For the latter, one defines a characteristic temperature,
\be\label{1.Teff}   T_{\rm eff}=\frac{\hbar c}{2a k_{\rm B}},
\ee
which, at room temperature,  corresponds to a separation $a\sim 4 \mu m$. The free energy depends typically on a dimensionless combination,
\be\label{1.T/Teff}   \frac{2a k_{\rm B}T}{\hbar c}=\frac{T}{T_{\rm eff}}.
\ee
It is, for most measurements, a small parameter making  the low temperature regime relevant.

At low temperature, the Matsubara sum in \Ref{1.F} becomes inefficient since high $l$ give significant contributions. Instead, one uses the Abel-Plana formula. For a sequence $f_l$ ($l$ integer) of numbers, the representation
\be\label{1.AP} \suml f_l=\int_0^\infty dl \,f(l)
                +\int_0^\infty \frac{dl}{e^{2\pi l}-1}\, i\left(f(il)-f(-i l)\right)
\ee
holds, where $f(l)$ is the analytic continuation to the complex $l$-plane with $f(l)=f_l$ for $l$ integer. Thereby it is assumed that the initial sum converges and that $f(l)$ does not have poles or branch points for $\Re l>0$. In case the derivatives exist, the right hand side of Eq. \Ref{1.AP} has an expansion,
\be\label{1.AP1} \suml f_l=\int_0^\infty dl \,f(l)
                +f'(0)+f'''(0)+\dots\,,
\ee
which, if applied to \Ref{1.F}, gives the expansion for $T\to0$.
Regrettably, in a number of cases, especially those related to the Drude model, these derivatives do not exist and a more elaborate treatment is needed. A  detailed discussion is given in the beginning of Sect. 3. It should be mentioned that \Ref{1.AP1} was used in the early calculations of the Casimir force, for instance in \cite{casi48-51-793}, to perform the limit of vanishing regularization parameter.

Using the Abel-Plana formula \Ref{1.AP}, the free energy \Ref{1.F} can be rewritten in the form
\be\label{1.1.F2}   \F=E_0+\Delta_T\F,
\ee
where
\be\label{1.E0} E_0=\frac{\hbar c}{4\pi^2}\sum_{\rm TE,TM}\int_0^\infty d\xi\, \phi(\xi),
\ee
resulting from the first integral in the right hand side of \Ref{1.AP}, is the vacuum energy, i.e., the zero temperature contribution, and the second integral,
\be\label{1.DF} \Delta_T\F =\frac{\hbar c}{4\pi^2} \int_0^\infty d\om\,\frac{1}{e^{\hbar\om/k_{\rm B}T}-1}\,
                \sum_{\rm TE,TM}\Phi(\om)
\ee
with
\be\label{1.Fa} \Phi(\om)=i\left(\phi(i\om)-\phi(-i\om)\right)
\ee
is the thermal contribution. We introduced the notation
\be\label{1.fi} \phi(\xi)=\int_0^\infty dk\,k\,\ln\left(1-r_1r_2 e^{-2a\eta}\right),
\ee
which will be used throughout this paper.
In \Ref{1.E0} the integration is over imaginary frequencies whereas in \Ref{1.DF} it is  over real frequencies and the integrand involves the Boltzmann factor $1/(e^{\hbar \om/k_{\rm B}T}-1)$.

Using representation \Ref{1.AP} or \Ref{1.AP1}, the low temperature expansion was calculated in nearly all cases of interest at least in the leading order. In the present paper we give a detailed representation of  this expansion in all cases of interest based on Eq. \Ref{1.DF}. We make full use of the analytic properties of the function $\phi(\xi)$, \Ref{1.fi}, and are able to improve a number of known expansions. Using this method we also reconsider the derivation of the contributions violating the third law of thermodynamics as well as the non-perturbative contribution appearing for vanishing relaxation parameter.

It is the aim of the present paper to review the low temperature expansion to the free energy for the basic models
and  represent them  unified  in the framework of Eq. \Ref{1.DF}. Thereby we consider the {\it asymptotic} expansion of the free energy as given by the Lifshitz formula, \Ref{1.LF}, for $T\to0$. This means that we do not consider any corrections which are exponentially small and we also do not discuss the applicability of the expansion to the one or other situation.  On the other side, we try to cover all relevant models and try to give a self contained representation which includes all derivations. It should enable the reader to follow all calculations nearly without consulting other sources.  Therefore, the  experienced reader may find some places too much going into detail for which the author asks for indulgences.
A part of the calculations is made   machinized, using a standard tool. We tried to explain all steps in such detail, that the reader should be able to repeat the calculations easily.

As for the models considered, these cover the most frequently used for the Casimir effect between real material bodies. We do not discuss their ranges of applicability. In this sense, the asymptotic expansions for low $T$ considered in this paper, must be understood primarily as properties of these models. Also, we do not consider all models. The model, describing a metal by impedance boundary conditions and the model, describing graphene by the Dirac model, are not considered here.

In the next section we collect the basic formulas for the free energy and, in a number of subsections, the specific models we are going to consider. These are, after  shortly recapitulating of the ideal conductor case, a   dielectric with fixed permittivity $\ep$ as simplest example for a medium. Next is the plasma model as the simplest model describing a metal beyond the ideal conductor. Then we add dissipation by considering a Drude model permittivity. The next subsection is devoted to an insulator, followed by a subsection for dielectric with dc conductivity as another example for dissipation. Finally we consider the hydrodynamic model for graphene.
In the third section we derive the low temperature expansion and the specific representation we are using. In section 4 we derive the low frequency expansions for the specific models. In the fifth section we collect the low temperature expansions for all models. Conclusions are given in  the last section. Some calculations are presented in the appendixes.

Following the theoretical approach of this paper we use units with $\hbar=c=k_B=1$.
Throughout the paper $\zeta_{\rm R}(s)$ denotes the Riemann zeta function and $\gamma_{\rm E}$ is Euler's constant.

\section{Basic formulas and models}
In this paper, the basic formula is the Lifshitz formula, mentioned already in the introduction. We consider two plane parallel interfaces perpendicular to the $z$-axis, with an empty gap of widths $a$ between them. On the interfaces, boundary or matching conditions are assumed to be given or the space behind the interfaces is assumed to be filled with a homogeneous medium or to be empty in case of graphene. Also we assume homogeneity and isotropy  in any plane parallel to the interfaces. We denote the Lifshitz formula in the form
\be\label{2.F} \F=\frac{T}{4\pi^2}\sum_{\rm TE,TM}\suml \phi(\xi_l)
\ee
with
\be\label{2.fi} \phi(\xi_l)=\int_0^\infty dk\,k\, \ln\left(1-r_1r_2 e^{-\eta}\right).
\ee
We allow for different reflection coefficients $r_1$ and $r_2$ on the two interfaces. One has to insert according to the model considered and to perform the summation over the polarizations in case of the \elm field.  In general, Eq. \Ref{2.F} is valid for any field if inserting the corresponding reflection coefficients in \Ref{2.fi}.
This formula represents the free energy $\F$ per unit area of an interface. The summation is over the Matsubara frequencies \Ref{1.xi} and for frequency dependent reflection coefficients $r_i(\om)$ one has to use their analytic continuation to $r_i(i\xi_l)$.

The Lifshitz formula was originally derived from the fluctuations of the \elm field in the gap introducing a random field into the Maxwell equations. At once it was mentioned that this field is associated with the 'zero point vibrations' \cite{lifs56-2-73}. Indeed, writing the vacuum energy \Ref{1.E0} in the form
\be\label{2.E0} E_0=\frac{1}{4\pi^2}\int_0^\infty d\xi\,\phi(\xi),
\ee
it is seen that this is just the half sum of the excitation of the \elm field in the sense introduced by Casimir in \cite{casi48-51-793} after Wick rotation. After that, the step from \Ref{2.E0} to \Ref{2.F} follows simply by applying the Matsubara formalism.

In \Ref{2.fi}, as compared with \Ref{1.fi}, we dropped a factor $2a$,
\be\label{2.drop}   e^{-2a\eta}\to e^{-\eta}.
\ee
This is equivalent to putting $a=1/2$ in all formulas. In fact, this is no restriction since this exponential is the only place where the separation $a$ enters the Lifshitz formula. Formally, this can be achieved by the substitution $k\to k/(2a)$. The dependence on $a$ can be restored at any time by dimensional consideration. Since the free energy is a density per unit area, its dependence is restored by
\be\label{2.restore}  \F\to\frac{1}{(2a)^3}\,\F
\ee
and all dimensional quantities entering $\F$ must be made dimensionless by the factor $2a$. For instance, one has to substitute the temperature by
\be\label{2.restore1}    T\to 2aT
\ee
and any frequency, like the plasma frequency, or dissipation parameter, $\ga$ or $\sigma$, by
\be\label{2.restore2}   \Om\to2a\Om.
\ee
In fact, there are no other parameter to be restored in this paper.
At this place we also mention how to restore the fundamental constants. For the free energy, in place of \Ref{2.restore}, one has to substitute
\be\label{2.restore3}   \F\to \frac{\hbar c}{(2a)^3}\,\F
\ee
(this is an energy divided by an area) and
\be\label{2.restore4}   T\to \frac{2ak_{\rm B}T}{\hbar c}, \quad \Om\to \frac{2a\Om}{c}
\ee
with $T$ to be measured in Kelvin and $\Om$ in $1/s$.

In the configuration of two parallel interfaces, considered in this paper, the polarizations of the \elm field always separate and are commonly chosen as transverse electric (TE) and transverse magnetic (TM) modes. The corresponding scalar amplitudes, $\Phi(t,\mathbf{x})$, satisfy the Maxwell equations and, on the interfaces, boundary or matching conditions. These amplitudes can be chosen as linear combinations of plane waves,
\be\label{2.mode}\Phi(t,\mathbf{x})\sim e^{-i\om t+i\mathbf{k}\mathbf{x}}
        \Phi(x_3),
\ee
with a dependence on the coordinate in perpendicular to the interfaces,
\be\label{2.mode3}\Phi(x_3)=\left\{\begin{array}{lc}
                                e^{ik_3x_3} &   (x_3<0\quad\mbox{or}\quad a<x_3),\\
                                e^{iqx_3}   &   (0<x_3<a ).
                              \end{array}
                        \right.
\ee
Here, $\omega$ is the frequency, $\mathbf{k}=\{k_1,k_2\}$ are the   wave vectors, resp. momenta (since we have $\hbar=1$), in directions parallel to the interfaces, and $k_3$ resp $q$, are the wave vectors in perpendicular direction outside, resp. inside, the gap. The dispersion relations,
\be  \label{2.disp1}\begin{array}{rcll}
\ep\om^2&=&k^2+k_3^2  ~~~~&\mbox{(outside the gap),}\\[4pt]
\om^2&=&k^2+q^2 &\mbox{(inside the gap),}
\end{array}
\ee
follow from inserting \Ref{2.mode} into the Maxwell equations. Here, the permittivity is allowed to depend on the frequency, $\ep=\ep(\om)$ and we use the notation $k=|\mathbf{k}|$. For non transparent boundary conditions like Dirichlet or ideal conductor conditions, the waves outside the gap do not contribute to the free energy \Ref{2.F}. For transparent boundary conditions like in the hydrodynamic model one has to put $\ep=1$ outside the gap and the corresponding momenta are equal, $k_3=q$.

After Wick rotation,
\be\label{2.Wick}\om  = i\xi,
\ee
one has always an imaginary wave vector
\be\label{2.q}q=i\eta
\ee
inside the gap. The wave vector $k_3$ outside the gap may remains real or it may become imaginary,
\be\label{2.}k_3=i\kappa.
\ee
The dispersion relations turn into
\be\label{2.disp2}\begin{array}{rcll}
                            \ep(i\xi)\xi^2&=&-k^2+\kappa^2  ~~~~&\mbox{(outside the gap),}\\
                            \xi^2&=&-k^2+\eta^2 &\mbox{(inside the gap).}
                    \end{array}
\ee
We will use  notations \Ref{2.q} - \Ref{2.disp2} throughout the paper.

In the Lifshitz formula, different choices of the integration variables are possible. In \Ref{2.F}, these are $k$ and $\xi$. In this case, one needs to express all other in terms of these,
\bea\label{2.disp3} \kappa&=&\sqrt{\ep(i\xi)\xi^2+k^2},    \nn\\
                    \eta&=&\sqrt{\xi^2+k^2}.
\eea
Below, the integration over $k$ in \Ref{2.fi} will be changed for $\eta$,
\be\label{2.fieta} \phi(\xi_l)=\int_\xi^\infty d\eta\,\eta\, \ln\left(1-r_1r_2 e^{-\eta}\right).
\ee
In that case one has to express
\bea\label{2.disp4} \kappa&=&\sqrt{(\ep(i\xi)-1)\xi^2+\eta^2},    \nn\\
                    k&=&\sqrt{\eta^2-\xi^2}
\eea
with the range $\eta\in [\xi,\infty)$.

As mentioned above, the properties of the interacting bodies enter the Lifshitz formula only through the reflection coefficients. For the interface at $x_3=a$, i.e., for the right one if looking on Fig. \ref{fig1}, the corresponding mode function is
\be\label{2.r2}    \Phi(x_3)=\left\{
            \begin{array}{lr}
                e^{iq(x_3-a)}+r \,e^{-i q (x_3-a)}& (x_3<a), \\
                t\, e^{ik_3(x_3-a)}                &   (x_3>a),
                    \end{array}     \right.
\ee
where $r$ and $t$ are the reflection and transmission coefficients. These are to be determined from the boundary or matching conditions in $x_3=a$. The reflection coefficient $r$, determined this way, must be inserted for $r_2$ in \Ref{2.fi}. The reflection coefficient $r_1$ follows, accordingly,  from the scattering from the right on the interface in $x_3=0$ with the choice
\be\label{2.r1}    \Phi(x_3)=\left\{
            \begin{array}{lr}
               t\, e^{-ik_3x_3}                &   (x_3<0), \\
                e^{-iqx_3}+r\, e^{i q x_3}& (0<x_3)

                    \end{array}     \right.
\ee
for the mode function. We mention that the reflection coefficients $r_1$ and $r_2$ are independent one from another. The Lifshitz formula stays correct for any combinations. In case of non-physical choices, like one from the TE and the other from the TM polarization, there would be, however,  no physical realization for.

As defined by Eqs.\Ref{2.r2} and \Ref{2.r1} together with Eq. \Ref{2.mode}, the reflection coefficients are functions of real $\om$, whereas $k_3$ and $q$ may be real or imaginary. For real $k_3$, the function $\Phi(x_3)$ describes scattering states and for imaginary both, $k_3$ and $q$, these are surface modes (for more details  see \cite{bord12-85-025005}).

In the remaining part of this section we specify the models which we will consider.

\subsection{Ideal conductor}
For ideal conducting surfaces, the boundary conditions are Dirichlet for the TE polarization and Neumann  for the TM polarization. The reflection coefficients are
\be\label{2.1.r}\rE=-1,\qquad r_{\rm TM}=1,
\ee
resulting in equal contributions from the two polarizations to the free energy. In this case, the mode functions $\Phi(x_3)$ terminate on the interfaces. Equivalently, one may put $t=0$ in \Ref{2.r2} and \Ref{2.r1}.
\subsection{Fixed permittivity}
For two dielectric half spaces with fixed permittivity $\ep$, the reflection coefficients are
\be\label{2.2.rq}\rE=\frac{q-k_3}{q+k_3},\quad \rM=\frac{\ep q-k_3}{\ep q+k_3}
\ee
in terms of the real wave numbers related by \Ref{2.disp1} with the frequency $\om$. In terms of the imaginary wave numbers we note
\be\label{2.2.reta}\rE=\frac{\eta-\kappa}{\eta+\kappa},\quad
                    \rM=\frac{\ep \eta-\kappa}{\ep \eta+\kappa},
\ee
and the relation \Ref{2.disp2} applies, including
\be\label{2.2.kappa}\kappa=\sqrt{(\ep-1)\xi^2+\eta^2}.
\ee
In this form, the reflection coefficients enter the Lifshitz formula \Ref{2.F} through \Ref{2.fieta}.

The reflection coefficients \Ref{2.1.r} of ideal conducting interfaces follow from the above with \Ref{2.2.rq} or \Ref{2.2.reta} in the limit $\ep\to\infty$. However, this does not imply that the free energy behaves the same way. This can be seen already in representation \Ref{2.F}. Consider $l=0$, i.e., the lowest contribution to the Matsubara sum. From Eq. \Ref{1.xi} we have $\xi_0=0$ and with \Ref{2.2.kappa} we note $\kappa=\eta$ in this case and the reflection coefficients become
\be\label{2.2.r0}\rE=0,\quad\rM=\frac{\ep-1}{\ep+1}\qquad (l=0).
\ee
Hence, in the limit $\ep\to\infty$, the $(l=0)$-contribution of the TE polarization does not deliver any contribution to the free energy while  all other contributions deliver the corresponding ideal conductor contributions. This behavior was first observed in \cite{schw78-115-1} and motivated 'Schwinger's prescription' to take the limit $\ep\to0$ before putting $l=0$. Also, the question on whether this single mode can influence the result much can be answered quite easily. In the high temperature limit, the $(l=0)$-contribution delivers the leading order contribution and with the vanishing TE contribution half of the result is missing. The physics behind this behavior is transparent. A fixed permittivity implies a dielectric material keeping its properties at all, including highest frequencies, which, of course, does not happen in physics.

\subsection{The plasma model}
The plasma model appears if one considers the whole space, or a half space behind an interface, being filled with a charged fluid (electrons, for example) coupled to the \elm field while the half space before the interface is empty. Eliminating the dynamical variables of the fluid from the equations of motion (or, in a functional integral approach, integrating them out), one comes to the same reflections coefficients \Ref{2.2.rq} or \Ref{2.2.reta} as above where one has to insert the  frequency dependent permittivity of the plasma model,
\be\label{2.3.ep}\ep^{\rm pl}(\om)=1-\frac{\Om^2}{\om^2}.
\ee
The causality of this permittivity was shown in \cite{klim07-40-F339}. Here $\Om$ is the so-called {\it plasma frequency} and from \Ref{2.disp4} or \Ref{2.2.kappa} we note
\be\label{2.3.kappa}\kappa=\sqrt{\Om^2+\eta^2}.
\ee
The spectrum, i.e., the mode content, of this model is well known. A recent discussion  in the context of vacuum energy was given in \cite{bord12-85-025005} and here we mention only that it is the same as for fixed permittivity with, in addition, a surface mode in the TM polarization. This is a mode with real frequency $\om$, propagating on the interface, and decaying exponentially on the vacuum side of the interface.

The plasma model describes some basic properties of the electrons in a metal. Typical values of the plasma frequency are approximately equal 8-9 eV. Its inverse is the skin depths
\be\label{2.3.skin}\delta=\frac{1}{\Om}.
 \ee
The reflection coefficients for ideal conducting boundary conditions can be obtained from \Ref{2.3.ep} in the limit $\Om\to\infty$. Also the free energy of ideal conductors is recovered in this limit  due to the sufficiently fast decrease of the permittivity for large frequencies.

\subsection{The Drude model}
This model is an extension of the plasma model allowing for dissipation. Physical reasons may be Ohmic losses or scattering of the electrons on the lattice or on impurities. These are accounted for by a phenomenological dissipation parameter $\ga>0$, entering the permittivity,
\be\label{2.4.ep}\ep^{\rm Dr}(\om)=1-\frac{\Om^2}{\om(\om+i\ga)},
\ee
of the model. In the formal limit $\ga\to0$ one recovers the permittivity $\ep^{\rm pl}(\om)$, Eq. \Ref{2.3.ep}, of the plasma model. The corresponding free energy does not follow in this limit, see Eq. \Ref{4.4.2.F}.

At the moment it is not clear whether the use of the Drude permittivity in the Lifshitz formula gives correct results or not \cite{BKMM,klim09-81-1827}. Since we do not enter this discussion in the present paper, we take this model as is and make only a few comments.

The permittivity \Ref{2.4.ep} is complex. Being inserted into the Maxwell equations \Ref{2.disp1}, a non vanishing imaginary part of the frequency results. For $\ga>0$, which one needs to assume, this describes dissipation of energy. This is in accordance with the intention of the model describing losses, finally resulting in heat. The model is in accordance with causality and the permittivity;  $\ep^{\rm Dr}(\om)$, Eq. \Ref{2.4.ep}, obeys the Kramers-Kronig relation. This model, taken alone, has no unitarity and a mode expansion of the free energy in terms of real frequencies is not possible \cite{bord11-71-1788}. In line with this, it must be mentioned  that after Wick rotation, which is possible for $\ga>0$, the permittivity
\be\label{2.4.ep2}\ep^{\rm Dr}(i\xi)=1+\frac{\Om^2}{\xi(\xi+\ga)}
\ee
is real delivering with \Ref{2.F} a real free energy. In this way one obtains an easy-to-use formula. Derivations of this procedure were given in \cite{bara75-18-305} and, recently discussed, for example,  in \cite{rosa10-81-033812}.

\subsection{Insulator described by oscillator model}
The response of insulators to the \elm excitations is, beyond a fixed permittivity,  frequently described by the permittivity
\be\label{2.5.ep}\ep^{\rm insul.}(\om)=1+\sum_{j=1}^{N}\frac{g_j}{\om_j^2-\om^2-i\om\ga_j}
\ee
of an $N$-oscillator model, where $\om_j$ are the oscillator frequencies, $g_j$ their strengths and $\ga_j>0$ their damping parameters. Here one excludes the case $\om_j=0$ since that would rather to be described by a plasma or Drude model. In the low frequency limit, $\om\to0$, one comes to a constant permittivity,
\be\label{2.5.ep1}\ep(0)=1+\sum_{j=1}^{N}\frac{g_j}{\om_j^2}\equiv \ep_0,
\ee
as considered in subsection 2.2.

It must be mentioned that \Ref{2.5.ep} includes also dissipation processes like the Drude model. However, the free energy calculated from Eq. \Ref{2.F} is real and for all non-vanishing oscillator frequencies, $\om_j\ne0$,  this model is not known to have problems \cite{BKMM}.

\subsection{The case of  dc conductivity}
At non zero temperature,   dielectrics posses, as a rule, some conductivity due to dissipation processes like in the Drude model. These are, in the simplest case, accounted for  by an additional contribution to the permittivity $\ep^{\rm insul.}(\om)$, Eq. \Ref{2.5.ep},
\be\label{2.6.ep}\ep^{\rm dc}(\om)=\ep^{\rm insul.}(\om)+i\frac{4\pi\sigma}{\om},
\ee
where $\sigma$ is the static conductivity resulting in a dc current. Being inserted into the Lifshitz formula, Eq. \Ref{2.F}, this model results in a real free energy. This conductivity is typically a function of temperature, $\sigma(T)$, vanishing at $T\to0$.   This model has problems similar to that in the Drude model mentioned in Subsection 2.4 \cite{BKMM}. Below we consider for completeness   also the case of a fixed $\sigma$, although it may be physically less interesting.

\subsection{Hydrodynamic model for graphene}
In the hydrodynamic model one assumes  a charged fluid, like in the plasma model, but confined to a plane, i.e., being two-dimensional. Again, eliminating the dynamical variables of the fluid, the Maxwell equations appear and the field strengths obey matching conditions on the pane. These result in reflection coefficients
\be\label{2.7.r}\rE=\frac{-1}{1-\frac{ i q}{\Om}},\quad
        \rM=\frac{1}{1+\frac{\om^2}{i q \Om}}.
\ee
In the mode expansion \Ref{2.r2} and \Ref{2.r1} one has to put $k_3=q$ since from both sides of an interface we have empty space. Accordingly, from the Maxwell equations \Ref{2.disp1}, only the second applies.

The matching condition for the TE mode is equivalent to a scalar field with a repulsive delta function potential on the interface obeying the wave equation
\be\label{2.7.delta}\left(-\frac{d^2}{dx_3^2}+2\Om\, \delta(x_3)\right)\Phi(x_3)=q^2\Phi(x_3).
\ee
For the TM mode, the corresponding scalar problem can be formulated in terms of a $\delta'$-potential.

The mode content of this model is quite similar to that of the plasma model considered in subsection 2.3. For instance, there is, for each interface, a surface mode. In the limit of infinite plasma frequency, $\Om\to\infty$, the reflection coefficients turn into that of an ideal conductor, Eq. \Ref{2.1.r}, and the free energy of this model turns into that of ideal conductors.

This model was first considered in \cite{fett73-81-367}. In  \cite{BIII} and subsequent papers it was used to describe the $\pi$-electrons of graphene and $C_{60}$. It provides a quite good description of their properties in interaction with \elm fields at large frequencies. For small frequencies the Dirac model \cite{seme84-53-2449,divi84-29-1685,kane05-95-146802}
provides a better description.

\section{Low temperature expansion for the free energy}
We take the Lifshitz formula in Matsubara representation, Eq.  \Ref{2.F}, as starting point for the low temperature expansion. The convergence of the sum in \Ref{2.F} and of the integration over $k$ in \Ref{2.fi} comes from the exponential factor
\be\label{3.1}e^{-2a\sqrt{\xi_l^2+k^2}}
\ee
(we restored, for a moment, the dependence on the gap's width $a$) making especially the sum over $l$ fast convergent. This picture changes with decreasing temperature $T$ since $l$ enters through the Matsubara frequency $\xi_l=2\pi T l$, Eq. \Ref{1.xi}. Obviously, for $\xi_l$ becoming large, numbers $l>1/T$ must be accounted for. Thus, for decreasing $T$, the convergence slows down and equation \Ref{2.F} becomes, in the limit, unusable.

A way out can be found if an analytic continuation of $\phi(\xi_l)$ to non-integer, in general complex, $\xi$  can be found. This gives the possibility to define $\phi(\xi)$, Eq. \Ref{2.fi}, as a function in the complex $\xi$-plane and, using the Cauchy theorem, to  represent the Matsubara sum in \Ref{2.F} as an integral,
\be\label{3.F1} T\suml\phi(\xi_l)=\frac{T}{2}\,\phi(0)+T\int_\Gamma  dl\,\frac{1}{1-e^{-i2\pi l}}\,
                                        \phi(2\pi T l).
\ee
Here the path $\Gamma$ encircles the non negative integers, $l=1,2,\dots$, and crosses the real axis in $l=\delta$ with $0<\delta<1$. The next step is a deformation of the integration path towards the imaginary axis. For $\Im l>0$, i.e., on the upper half of the path, one substitutes
\be\label{3.l+}   l=\frac{i\om}{2\pi T}
\ee
with $\om\in[0,\infty)$ and the exponential in the denominator becomes large for $\om\to\infty$. For $\Im l<0$, i.e., on the lower half of the path, one substitutes
\be\label{3.l-}l=\frac{-i\om}{2\pi T}.
\ee
Since, in this case,  the exponential does not grow for large $\om$, one needs to rewrite it,
\be\label{3.tr}\frac{1}{1-e^{-i2\pi l}}=1-\frac{1}{1-e^{i2\pi l}}.
\ee
In the contribution from the first term on the right hand side it is meaningful to change the integration variable according to $l=\frac{\xi}{2\pi T}$ and to write down this contribution separately. The integration can go along the real axis since there are no poles in this contribution. The second term can be joined with the contribution from the upper half of the path. In both cases $\om$ runs from zero till infinity.
One comes to the representation
\be\label{3.F2}T\suml\phi(\xi_l)=\frac{1}{2\pi}\int_0^\infty d\xi\,\phi(\xi)
        +\frac{1}{2\pi}\int_0^\infty d\om \, \frac{1}{e^{\om/T}-1}
            i\left(\phi(i\om)-\phi(-i\om)\right).
\ee
This is the well known {\it Abel-Plana formula}.

In moving the integration path $\Gamma$ towards the imaginary axis and performing the limit $\delta\to0$, from the origin, i.e., from $l=0$, a contribution appeared which just cancels the first term in the right hand side of Eq. \Ref{3.F1}. We mention that in Eq. \Ref{3.F2} there is no pole for $\om=0$ due to the compensation in the parentheses. Further we mention, that in \Ref{3.F2} it is assumed that the function $\phi(\xi)$ is continuous in $\xi=0$. If this is not the case, one cannot move the path completely to the imaginary axis. However, such situation does not appear in the examples considered in this paper.

A further assumption in deriving Eq. \Ref{3.F2} concerns the function $\phi(\xi)$. It is assumed that it does not have poles or branch points in the half plane $\Re \xi>0$. Otherwise, from moving the path there would be additional contributions. This property is always guaranteed if the modes of the \elm field are subject to an elliptic scattering problem. It holds also for the model with dissipation where the corresponding poles are all located in the half plane $\Re \xi<0$. For vanishing dissipation parameter, these move towards the imaginary axis from the left and the path must pass them on the right side, for instance, by adding an infinitesimal amount,
\be\label{3.+0}\phi(\pm i \om)\to\phi(\pm i \om+0),
\ee
which is necessary for all models anyway.

In application to the free energy  \Ref{2.F}, Eq. \Ref{3.F2} defines a split,
\be\label{3.F3} \F=E_0+\Delta_T\F,
\ee
into the vacuum energy $E_0$ \Ref{2.E0},
\be\label{3.E0}E_0=\frac{1}{4\pi^2}\int_0^\infty d\xi\,\phi(\xi),
\ee
resulting from the first term in the right hand side of \Ref{3.F2}, and, from the second term, the temperature dependent part,
\be\label{3.F4}\Delta_T\F=\frac{1}{4\pi^2}\int_0^\infty d\om \, \frac{1}{e^{\om/T}-1} \, \Phi(\om),
\ee
involving the Boltzmann factor $1/(e^{\om/T}-1)$ and
\be\label{3.Fi}\Phi(\om)= i\left(\phi(i\om)-\phi(-i\om)\right).
\ee
The integration variable $\om$ has the meaning of a frequency like that entering Eq. \Ref{2.disp1} provided a mode expansion makes sense. As already mentioned, this is not the case for models with dissipation (see the remark at the end of this section). However, independently on the interpretation, representation \Ref{3.F3} with \Ref{3.F4}  and the property \Ref{3.Fi} are valid for these too.

At this place, an important remark on the direction of the contour rotations is in order. The rotation \Ref{3.l-} is the inverse of the usual Wick rotation \Ref{2.Wick}, whereas \Ref{3.l+} is the inverse of an Anti-Wick rotation. Since it is customary to write the Abel-Plana formula just with the order of terms as in the parentheses in \Ref{3.F2} with $\phi(i\om)$ first, in application to the free energy \Ref{3.Fi}, the term corresponding to the inverse Anti-Wick rotation, goes first. Of course, this does not change anything except for notations.

Below we will find it convenient, in a number of occasions, especially after some variable substitutions, to use the reflection property,
\be\label{3.refl}\phi(-i\om)=\phi(i\om)^*,
\ee
this function  with reflection coefficients following from a scattering problem has, to represent the difference in \Ref{3.Fi} in the form
\be\label{3.cc}i(\phi(i\om)-\phi(-i\om))=-(\phi(i\om)-{\rm c.c.}),
\ee
where c.c. denotes the complex conjugate of what is in front to be inserted. In doing so one has only to pay attention to signs in some places, especially in the permittivity, which changes under Wick rotation $\ep(\om)\to\ep(i\xi)$, but which enters $\phi(i\om)$ after Anti-Wick rotation,
\be\label{3.AW}\ep(i\xi)\to\ep(-\om) .
\ee
This sign shows up in models with dissipation only.

In some simple cases it is possible to use the Abel-Plana formula, formally not entering the complex plane. Assume the function $\phi(\xi)$ has a Taylor series expansion,
\be\label{3.Tay}\phi(\xi)=\sum_{n\ge 0}\frac{\xi^n}{n!}\,\phi^{(n)}(0),
\ee
one gets for $\Phi(\om)$, Eq. \Ref{3.Fi}, an expansion directly in terms of real quantities. This can be inserted into \Ref{3.F4}. Interchanging the orders of integration and summation, the integration can be carried out. One obtains
\be\label{3.F5} \Delta_T\F=\frac{1}{4\pi^2}\sum_{k\ge 0}(-1)^{k+1}{2\zeta_{\rm R}(2k+2)} \,\phi^{(2k+1)}(0)\,T^{2k+2},
\ee
which is known as Euler-Maclaurin summation formula,
with the Riemann zeta function in even integers,
\be\label{3.zetaR}\zeta_{\rm R}(2k+2)=(-1)^{k+1} \frac{(2\pi)^{2k+2}B_{2k+2}}{2(2k+2)!},
\ee
in terms of the Bernoulli numbers $B_n$. Obviously, this is an expansion for $T\to0$. For most systems, however, the function $\phi(\xi)$ does not have a Taylor expansion. Typically, in the examples considered in this paper,  $\phi(0)$ exists, but not the derivatives. Nevertheless, even in the case there is no Taylor expansion, the derivatives, as far as they exist, give with Eq. \Ref{3.F5} the lowest contributions to the asymptotic expansion for $T\to0$.

We add a remark on the convergence of the vacuum and the free energies. In general, the vacuum energy, and with it also the free energy, have ultraviolet divergences resulting from slow convergence for large frequencies or momenta. In the situation of a Casimir effect setup, considered here, these divergences do not depend on the width of the gap and the Casimir force is always finite. The split \Ref{3.F3} is, of course,  valid beyond the Casimir effect setup. In that case the divergences in the free and in the vacuum energies are the same and the thermal part $\Delta_T\F$, \Ref{3.F4}, does not have any divergencies. Its convergence follows, obviously, from the Boltzmann factor, whereas before the contour rotation, i.e., in Eq. \Ref{3.F1}, this factor is bounded. It is just the contour rotation, which, without changing the integral, redistributes for small $T$ the main contribution to the integral towards small $\om$. As a result, for $T\to0$, the integral over $\om$ is fast convergent and the contributions from $\om\gtrsim 0$ determine the asymptotic expansion for low $T$. This is in opposite to the situation in the initial Matsubara representation where large imaginary frequencies $\xi_l$ were needed for.

At this place we mention the Poisson re-summation  formula which is yet another way to redistribute the convergence. In order to use that formula either one has an explicit representation, typically an Gaussian exponential, or one needs to make an analytic continuation from integer $l$ to, at least, real ones. One obtains, in place of the Matsubara sum, another sum, which is fast converging for small $T$. We do not use this approach in the present paper.

Due to the convergence properties just discussed, Eq. \Ref{3.F4} allows, in a simple way, for the low temperature expansion of $\Delta_T\F$. For this, it is sufficient to assume the function $\Phi(\om)$, Eq. \Ref{3.Fi}, has an asymptotic expansion for $\om\to0$,
\be\label{3.Fi2}\Phi(\om)=\left(\Phi_1+\tilde{\Phi}_1 \ln \om\right)\om
                +\Phi_{\frac32} \om^{3/2}
                +\Phi_2 \om^2
                +\Phi_{\frac52} \om^{5/2}
                +\Phi_3 \om^3
                +\dots\,.
\ee
Here we allowed for a logarithmic contribution in the first order and for half-integer orders since these will appear below in the  Drude model (section 4.4) and for the insulator (section 4.5).   If the low frequency expansion \Ref{3.Fi2} is found, the asymptotic expansion, for $T\to0$, of the free energy can be easily written down by inserting \Ref{3.Fi2} into \Ref{3.F4} and
using
\be\label{3.zeta} \int_0^\infty d\om\frac{\om^s}{e^{\om/T}-1}=\Gamma(s+1)\zeta_{\rm R}(s+1)T^{s+1}
\ee
for the integration over $\om$. For logarithmic contributions one may take the derivative of this formula with respect to $s$. In this way one comes to
\bea\label{3.F6}\Delta_T\F&=&\frac{1}{4\pi^2}
        \left[
       \left(\left(\Phi_1+({\zeta_{\rm R}'(-1)}+\ln(2\pi T)\right)\tilde{\Phi}_1\right)\zeta_{\rm R}(2)T^2
       +\frac{3\sqrt{\pi}\zeta_{\rm R}(5/2)}{4}\Phi_{\frac32}T^{5/2}
       \right.
\nn\\ & &\left.~~~~
        +2\zeta_{\rm R}(3)\Phi_2 T^3
         +\frac{15\sqrt{\pi}\zeta_{\rm R}(7/2)}{8}\Phi_{\frac52}T^{7/2}
         +6\zeta_{\rm R}(4)\Phi_3 T^4+\dots\right].
\eea
By using the explicit values \Ref{3.zetaR} and multiplying out the square bracket, this formula can be rewritten,
\bea\label{3.F6a}\Delta_T\F&=&
         \frac{1}{24} \left(\Phi_1+({\zeta_{\rm R}'(-1)}+
                \ln(2\pi T))\tilde{\Phi}_1\right)T^2
         +\frac{9\zeta_{\rm R}(5/2)}{16\pi^{3/2}}\Phi_{\frac32}T^{5/2}
\nn\\&&        +\frac{\zeta_{\rm R}(3)}{2\pi^2}\,\Phi_2 T^3
        +\frac{15\zeta_{\rm R}(7/2)}{32\pi^{3/2}}\Phi_{\frac52}T^{7/2}
        +\frac{\pi^2}{60}\,\Phi_3 T^4
        +\dots\,.
\eea
It should be mentioned that expansion \Ref{3.F6} can be derived for any model if $\Phi(\om)$ does not depend on $T$. It starts always at least from  $T^2$.
For this reason, it cannot come  in contradiction with thermodynamics.
However, in case of the Drude model with a dissipation parameter $\ga(T)$ vanishing for $T\to0$, the expansion \Ref{3.F6} is incomplete as discussed in detail in Sect. 4.4.2.
It should be mentioned that expansion \Ref{3.F5} is a special case of \Ref{3.F6} since it has only odd powers of $T$ and no logarithmic contributions.

\begin{figure}\unitlength1cm
\begin{picture}(8,6)
  \put(0,-3){\includegraphics[width=15 cm]{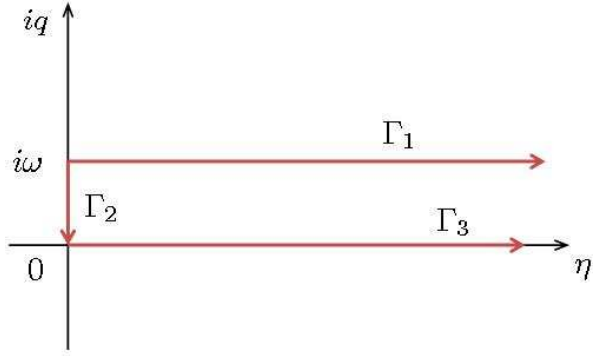}    }  \end{picture}
 \caption{The complex  plane of the wave number $\eta$, integration pathes shown. $\Gamma_1$ is the original path in Eq. \Ref{3.fi}. It is equivalent to $\Gamma_1$ (region \a) and $\Gamma_2$ (region\b) in \Ref{3.reg}. }\label{fig3}
\end{figure}

With representation \Ref{3.F6} resp. \Ref{3.F6a} at hand, the 'remaining task' is the calculation of the coefficients $\Phi_i$. For this we return to Eq. \Ref{2.fi},
\be\label{3.fi} \phi(\xi)=\int_0^\infty dk\,k\, \ln\left(1-r_1r_2 e^{-\eta}\right),
\ee
assuming the analytic continuation to the complex $\xi$-plane is done.
Following from Eqs. \Ref{2.disp1} and \Ref{2.disp2}, the variables $\om$, $k$ and $\eta$ are related by
\be\label{3.eta}\eta=\sqrt{k^2-\om^2}.
\ee
It is possible to change the integration in \Ref{3.fi} from $k$ to $\eta$, which runs from $\eta=i\om$ till infinity parallel to the real axis. In Fig. \ref{fig3} this integration path is denoted by $\Gamma_1$. In deriving the expansion for the various models in the next section, it turns out to be convenient to change the integration path for the sum of two, one running from $\eta=i\om$ to $\eta=0$, and the other, from $\eta=0$ along the real axis till infinity. These two are shown in Fig. \ref{fig3} as $\Gamma_2$ and $\Gamma_3$.  Of course, the integral does not change. In representation \Ref{3.fi} with integration over $k$, this corresponds to a subdivision of the integration region into two regions,
\be\label{3.reg}\begin{array}{ccll}
                  \mbox{region \a:}&0\le k\le\om,&\mbox{  with  }q=\sqrt{\om^2-k^2},\quad&\mbox{(scattering states)} \\
                   \mbox{region \b:}&\om\le k,&\mbox{  with  }\eta=\sqrt{k^2-\om^2},\quad&\mbox{(surface modes)}
                \end{array}
\ee
where, at once,  that wave numbers are shown which are real in the given region, as it follows from Eqs. \Ref{2.disp1} and \Ref{2.disp2}.
As a convention, we will all quantities calculated in these regions,  denote correspondingly by subscripts $(a)$ or $(b)$.

According to \Ref{3.reg}, the integral in \Ref{3.fi} splits into two,
\be\label{3.fi12}\phi(i\om)=\phi_\a (i\om)+\phi_\b(i\om),
\ee
which will be treated separately. With Eq. \Ref{3.Fi}, this induces a corresponding split
\be\label{3.Fi12}\Phi(\om)=\Phi_\a (\om)+\Phi_\b(\om),
\ee
and the relations
\bea\label{3.zush}  \Phi_\a (\om)&=&i(\phi_\a (i\om)-\phi_\a (-i\om)),
\nn\\               \Phi_\b (\om)&=&i(\phi_\b (i\om)-\phi_\b (-i\om)).
\eea
In region \a\, we have
\be\label{3.f11}\phi_\a (i\om)=\int_0^\om dk\,k\,\ln\left(1-r_1r_2e^{-iq}\right)
\ee
with $q$ shown in \Ref{3.reg}. The reflection coefficients must be expressed in terms of $\om$ and $k$. For instance, for the wave number $k_3$ we note
\be\label{3.k3}k_3=\sqrt{\ep\om^2-k^2},
\ee
which may be both, imaginary or real in dependence on the model.
Changing for the integration variable $q$, Eq. \Ref{3.f11} can be written in the form
\be\label{3.fq}\phi_\a (i\om)=\int_0^\om dq\,q\,\ln\left(1-r_1r_2e^{-iq}\right),
\ee
where one needs to express everything in terms of $q$ and $\om$, for instance $k=\sqrt{\om^2-q^2}$ and $k_3=\sqrt{(\ep-1)\om^2+q^2}$. In this formula, the $r_i$ are the reflection coefficients for scattering states.

In region \b\, we have
\bea \label{3.f22}\phi_\b (i\om)&=&\int_\om^\infty dk\,k\,\ln\left(1-r_1r_2e^{-\eta}\right)
 \nn   \\      &=& \int_0^\infty d\eta\,\eta\,\ln\left(1-r_1r_2e^{-\eta}\right).
\eea
In the second line we changed the integration variable for $\eta$ using \Ref{3.reg}. This integration corresponds to the path $\Gamma_3$ in Fig. \ref{fig3}. The reflection coefficients entering the second line must be expressed in terms of $\om$ and $\eta$. For instance, from \Ref{2.disp2} we note
\be\label{3.kappa}\kappa=\sqrt{-(\ep-1)\om^2+\eta^2}.
\ee
In \Ref{3.fq}, the $r_i$ are the reflection coefficients analytically continued into region $\b$.
We remind that we keep the relations $\kappa=ik_3$ and $\eta=iq$ in all calculation.

\section{The low frequency expansion for specific models}
In this section we obtain the  low frequency  expansions \Ref{3.Fi2} of the function $\Phi(\om)$ for various models. This section comprises the main technical part of the paper. Some calculations are banned to the appendixes. As mentioned in the Introduction, we use the simplest form of notations, especially we drop the factor $2a$ everywhere as announced.

\subsection{Ideal conductor}
This is the simplest model and well known. We consider it for completeness. At once we illustrate the technique used, especially the division of the integration in $\phi(i\om)$ into two regions. Also it allows for an easy checking of the overall factors.

The reflection coefficients for ideal conductors are given by Eqs. \Ref{2.1.r} and their product, entering $\phi(i\om)$, is $r_1r_2=1$ for both polarizations. The contribution from region \a\, can be written in the form
\be\label{4.1.fi}\phi_\a(i\om)=\int_0^\om dq\,q\,\ln\left(1-e^{-iq}\right),
\ee
where we changed the integration variable for $q$ using \Ref{3.reg}. The logarithm can be written in the form
\be\label{4.1.ln}\ln\left(1-e^{-iq}\right)=-\frac{iq}{2}+i\frac{\pi}{2}+
\ln\left(2\sin\frac{q}{2}\right).
\ee
Now, as long as $\om<\pi$, the sine does not change sign and the and the logarithm in the right side stays real. Now we calculate according to \Ref{3.zush} the imaginary part,
\bea\label{4.1.Fi1}\Phi_\a(\om)&=&i(\phi_\a(i\om)-c.c.), \nn\\
                                &=&\int_0^\infty dq\,q\,(q-\pi).
\eea
The remaining integration is trivial,
\be\label{4.1.Fi2}\Phi_\a(\om)=-\frac{\pi}{2}\om^2+\frac{1}{3}\om^3.
\ee
In region \b\, we note
\be\label{4.1.fi2}\phi_\b(i\om)=\int_0^\infty d\eta\,\eta\,\ln\left(1-e^{-\eta}\right).
\ee
This expression is completely real. Hence it does not contribute, $\Phi_\b(\om)=0$, and we get from \Ref{3.Fi12} and \Ref{4.1.Fi2}
\be\label{4.1.Fi}\Phi(\om)=-\frac{\pi}{2}\om^2+\frac{1}{3}\om^3.
\ee
We mention that this formula not only provides the asymptotic expansion for $\om\to0$, it is exact for $\om<\pi$.

From Eq. \Ref{4.1.Fi}, in the context of Eq. \Ref{3.Fi2}, we have non vanishing coefficients
\be\label{4.1.Fcoeffs}\Phi_2=-\frac{\pi}{2},\quad \Phi_3=\frac{1}{3}.
\ee
Inserted into Eq. \Ref{3.F6a}, these deliver the expansion for ideally conducting interfaces,
\be\label{4.1.DF1}\Delta_T\F^{\rm id}=-\frac{\zeta_{\rm R}(3)}{2\pi}\,T^3+\frac{\pi^2}{90}\,T^4+\dots\,.
\ee
Here we included a factor of 2 to account for the two polarizations of the \elm field.
The dots represent exponentially decreasing contributions we do not care of in the present paper.

It should be mentioned that the method used in this paper is only one out of quite a number of equivalent ones applicable for this simple model.  More details can be found, for example, in \cite{BKMM}, chap. 7.4.

\subsection{Fixed permittivity}
Dielectrics with fixed permittivity $\ep$ represent the simplest model for an insulator. While the ideal conductor considered in the preceding subsection is in the aim of Casimir's original idea, an insulator is rather in the spirit of Lifshitz's approach. Here both are treated within the same formalism. Also, the model with a fixed $\ep$  may serve as a good approximation for more complicated permittivities at low frequency.

The reflection coefficients are given by Eqs. \Ref{2.2.rq}, or by \Ref{2.2.reta}, which will be used in regions 1 and 2, accordingly. As already mentioned, the limit $\ep\to\infty$ does not turn the free energy into that of ideal conducting interfaces. Hence, a situation with one interface ideal conducting, the other with finite permittivity $\ep$ behind, is different from a situation with both interfaces having finite permittivities $\ep_1$ and $\ep_2$ behind and cannot be obtained by any limiting process. For this reason we consider 4 cases as shown in Table \ref{4.2.cases} and denote the case as an index in parentheses, $\Phi_{(k)}(\om)$ (k=1,...,4).
\begin{table}[h]
  \centering
 \begin{tabular}{|c|c |c|c|}
  \hline
  case &first interface&second interface &polarization\\
   & $r_1$ & $r_2$ & \\ \hline
  1  &  $-1$ & $r_{\rm TE}$ & TE \\
  2  & $r_{\rm TE}$  & $r_{\rm TE}$  & TE \\
  3  &   $1$  & $r_{\rm TM}$  & TM \\
  4  & $r_{\rm TM}$  & $r_{\rm TM}$  & TM \\
  \hline
\end{tabular}
  \caption{Notations for the four cases considered.}\label{4.2.cases}
\end{table}

As discussed in Sect. 3 we will perform the calculations separately in regions \a\, and \b. We add this information to the index such that
\be\label{4.2.ki}\Phi_{(k,n)}(\om)\qquad (k=1,...,4,\ \ n=a,b)
\ee
denotes the contribution from region $n$ to case $k$. We  use his notation also for the functions $\phi(i\om)$ and in the relations
\be\label{4.2.Fifi}\Phi_{(k,n)}(\om)=i\left(\phi_{(k,n)}(i\om)-c.c.\right).
\ee
For a given case, the contributions from both regions must be added,
\be\label{4.2.12}\Phi_{(k)}(\om)=\Phi_{(k,a)}(\om)+\Phi_{(k,b)}(\om).
\ee
In the final result for the \elm field, cases 1 and 3 or cases 2 and 4 must be added.

\subsubsection{Region \a}
Here we use the reflection coefficients as given by Eq. \Ref{2.2.rq}. In Eq. \Ref{3.f11} we change, for convenience, the integration variable for $q=\sqrt{\om^2-k^2}$. In this case, in Eq. \Ref{2.2.rq}, one has to use $k_3=\sqrt{(\ep-1)\om^2+q^2}$ and we get
\be\label{4.2.1.fi}\phi_{(k,a)}(i \om)=
        \int_0^\om dq\,q\,\ln\left(1-r_1r_2e^{-iq}\right),
\ee
where for $r_1$ and $r_2$ one has to insert according to Table \ref{4.2.cases}.

In this expression, a direct expansion of the integrand in powers of $\om$ with subsequent integration over $q$ delivers the expansion of $\phi_{(k,a)}(i \om)$ and, by means of Eq. \Ref{4.2.Fifi}, that of $\Phi_{(k,a)}( \om)$. In the latter only odd powers of $\om$ remain. Defining expansion coefficients $\Phi_{(k,a),i}( \om)$ in parallel to Eq. \Ref{3.Fi2}, these can be calculated easily by machine. The   non vanishing coefficients are shown up to the order $i=6$, which corresponds to an order $T^7$ in the expansion \Ref{3.F6a},
\begin{eqnarray}\label{4.2.1.Fii}
\Phi_{(1,a),3}&=&  \frac{1}{3} \left(-{\ep}^{3/2}+({\ep}-1)^{3/2}+1\right), \\ \nn
\Phi_{(1,a),5}&=&  \frac{{\ep}-1}{36}  \left({\ep}^{3/2}-({\ep}-1)^{3/2}\right), \\ \nn
\Phi_{(2,a),3}&=&  \frac{\sqrt{{\ep}}-1}{6} \left(\left(\sqrt{{\ep}}+1\right) \left(\sqrt{{\ep}}-\sqrt{{\ep}-1}\right)-2\right) , \\ \nn
\Phi_{(2,a),5}&=&  \frac{({\ep}-1)^2}{96 \sqrt{{\ep}}}\left(3\sqrt{{\ep}} \left(\sqrt{{\ep}}-\sqrt{{\ep}-1}\right)-1\right) , \\ \nn
\Phi_{(3,a),3}&=&  \frac{\sqrt{{\ep}}-1}{3}  \left(\left(1-2 \left(\sqrt{{\ep}}-\sqrt{{\ep}-1}\right) \sqrt{{\ep}-1}\right) \left(\sqrt{{\ep}}+1\right) \sqrt{{\ep}}-1\right), \\ \nn
\Phi_{(3,a),5}&=&  \frac{({\ep}-1) {\ep}}{180}  \left(\sqrt{{\ep}} \left(48 {\ep}^3-72 {\ep}^2+10 {\ep}+15\right)-8 ({\ep}-1)^{3/2} \left(6{\ep}^2-1\right)\right), \\ \nn
\Phi_{(4,a),3}&=&  \frac{\left(\sqrt{{\ep}}-1\right)^{3/2}}{6 {\ep}} \left({\ep}\sqrt{\sqrt{{\ep}}-1} \left(2 {\ep}^{3/2}+4 {\ep}+3 \sqrt{{\ep}}+2\right)-\left(\sqrt{{\ep}}+1\right)^{3/2} \left(2 {\ep}^2-1\right)\right), \\ \nn
\Phi_{(4,a),5}&=&  \frac{({\ep}-1)^2 }{1440 {\ep}^3}\left({\ep}^{5/2}(2 {\ep}-1) \left(24 {\ep}^3-7 {\ep}-3\right)+\sqrt{{\ep}-1} \left(-48 {\ep}^6+8 {\ep}^4-2 {\ep}^2-3\right)\right).
\end{eqnarray}

\subsubsection{Region \b}
In this region we use Eq. \Ref{3.f22} and insert the reflection coefficients as given by Eq. \Ref{2.2.reta} with $\kappa=\sqrt{\eta^2-(\ep-1)\om^2}$, which follows from Eq. \Ref{2.disp4}. Next we observe that in
\be\label{4.2.2.fi}\phi_{(k,b)}(i \om)=\int_0^\infty d\eta\,\eta\,\ln\left(1-r_1r_2e^{-\eta}\right)
\ee
the part of the integration with $\eta>\sqrt{\ep-1}\,\om$ delivers a real contribution and does not contribute in \Ref{4.2.Fifi}. Thus we restrict the integration region,
\be\label{4.2.2.F1}\Phi_{ (k,b)}(\om)=\int_0^{\sqrt{\ep-1}\om}d\eta\,\eta\,
                    i\left(\ln\left(1-r_1r_2e^{-\eta}\right)-{\rm c.c.}\right).
\ee
In this expression $\kappa$ is imaginary, $\kappa=ik_3$ with $k_3=\sqrt{(\ep-1)\om^2-\eta^2}$. Now we make the substitution $\eta\to\sqrt{\ep-1}\,\om\,\eta$ and get
\be\label{4.2.2.F2}\Phi_{ (k,b)}(\om)=
            (\ep-1)\om^2\,\Psi\left(\frac{\sqrt{\ep-1}\,\om}{2}\right)
\ee
with
\be\label{4.2.2.psi}\Psi(\mu)=\int_0^1 d\eta\,\eta\,
            i\left(\ln\left(1-r_1r_2e^{-2\mu\eta}\right)-{\rm c.c.}\right)
\ee
and for the reflection coefficients, in dependence on the polarization, one has here to insert
\be\label{4.2.2.rX}\rE=\frac{\eta-i\sqrt{1-\eta^2}}{\eta+i\sqrt{1-\eta^2}}, \qquad
            \rM=\frac{\ep\eta-i\sqrt{1-\eta^2}}{\ep\eta+i\sqrt{1-\eta^2}}.
\ee
For the expansion of the function $\Phi_{ (k,2)}(\om)$, Eq. \Ref{4.2.2.Fi}, for $\om\to0$, we need to know the expansion of the function $\Psi(\mu)$ for $\mu\to0$.

It is not possible to expand the integrand in \Ref{4.2.2.psi} since a subsequent integration would not converge for $\eta\to1$ since for the reflection coefficients \Ref{4.2.2.rX} the relations
\be\label{4.2.2.r1}\lim_{\eta\to1}\rE=1, \qquad \lim_{\eta\to1}\rM=1
\ee
hold, which result in singular terms in an expansion of the logarithm in \Ref{4.2.2.psi}. We proceed by factorizing the logarithm in \Ref{4.2.2.psi} and representing $\Psi(\mu)$ as a sum,
\be\label{4.2.2.AB}\Psi(\mu)=\Psi_A(r_1r_2,\mu)+\Psi_B(r_1r_2,\mu)
\ee
of two functions,
\bea\label{4.2.2.A,B}\Psi_A(r,\mu)&=&\int_0^1 d\eta\,\eta\,
            i\left(\ln\left(1+re^{-\mu\eta}\right)-{\rm c.c.}\right),
\nn\\       \Psi_B(r,\mu)&=&\int_0^1 d\eta\,\eta\,
            i\left(\ln\left(1-re^{-\mu\eta}\right)-{\rm c.c.}\right).
\eea
The functions $\Phi_{(k,b)}(\om)$, defined in \Ref{4.2.ki}, can be reobtained from from these as follows,
\be\label{4.2.2.Fic}\begin{array}{llll}
    \mbox{case 1:} & \Phi_{(1,b)}(\om)&=& (\ep-1)\om^2\,\Psi_A(1,\sqrt{\ep-1}\,\om),
\\[10pt]  \mbox{case 2:} & \Phi_{(2,b)}(\om)&=& (\ep-1)\om^2\,
                \left(\Psi_A(1,\frac12\sqrt{\ep-1}\,\om)
                    +\Psi_B(1,\frac12\sqrt{\ep-1}\,\om)\right),
\\[10pt]  \mbox{case 3:} & \Phi_{(3,b)}(\om)&=&
                            (\ep-1)\om^2\,\Psi_B(\ep,\sqrt{\ep-1}\,\om),
\\[10pt]  \mbox{case 4:} & \Phi_{(4,b)}(\om)&=& (\ep-1)\om^2\,
                \left(\Psi_A(\ep,\frac12\sqrt{\ep-1}\,\om)
                    +\Psi_B(\ep,\frac12\sqrt{\ep-1}\,\om)\right),
                    \end{array}
\ee
where we used ${\rM}_{|\ep=1}=\rE$, which holds for the coefficients \Ref{4.2.2.rX}.

For the function $\Psi_A(\mu)$, the expansion can be obtained easily by expanding the integrand with subsequent integration. The singularities resulting from the expansion of the logarithm appear for $\eta=0$ and are compensated by factors from the expansion of the exponential. We denote this expansion in the form
\be\label{4.2.2.Ai}\Psi_A(\mu)=\sum_{i\ge 0}w_{A,i}(\ep)\mu^i.
\ee
The coefficients up to order 4 are shown in Table \ref{4.2.w}.

For the function $\Psi_B(\mu)$, which carries the above mentioned singularities if expanding the integrand, we define an auxiliary function
\be\label{4.2.2.aux}\Psi_{\rm aux}(\mu)=
        \int_0^1 d\eta\,\eta\,i \left(
        \ln\left(1-\frac{\ep-i\sqrt{1-\eta^2}}{\ep+i\sqrt{1-\eta^2}}\,e^{-\mu}\right)
        -{\rm c.c.}\right).
\ee
In this function, the integration can be carried out explicitly with subsequent expansion in powers of $\mu$. We use this function to represent $\Psi_B(\mu)$ in the form
\be\label{4.2.2.psiB}\Psi_B(\mu)=\Psi_{\rm sub}(\mu)+\Psi_{\rm aux}(\mu)
\ee
with
\be\label{4.2.2.psisub}\Psi_{\rm sub}(\mu)=
    \int_0^1 d\eta\,\eta\,i \left(
     \ln\left(1-r\,e^{-\mu\eta}\right)-
        \ln\left(1-\frac{\ep-i\sqrt{1-\eta^2}}{\ep+i\sqrt{1-\eta^2}}\,e^{-\mu}\right)
        -{\rm c.c.}\right).
\ee
In this function, the integrand can be expanded up to the order of $\mu^4$ with convergent subsequent integration over $\eta$. In this way we obtain the expansion
\be\label{4.2.2.Bi}\Psi_B(\mu)=\sum_{i= 0}^{4}w_{B,i}(\ep)\mu^i+\dots.
\ee
All these operations can be carried out by machine. The coefficients up to order 4 are shown in Table \ref{4.2.w}.

\begin{table}[h]
  \centering$
  \begin{array}{c|ccccc}
 i&0&1&2&3&4  \\ \hline
 w_{A,i}(\ep)&  \frac{\pi }{2 (\ep+1)} & -\frac{1}{3 \ep} & 0 & \frac{2 \ep^2+3}{180
   \ep^3} & 0 \\[8pt]
w_{B,i}(\ep)& -\frac{\pi  \ep}{2 (\ep+1)} & \frac{2 \ep}{3} & -\frac{\pi
   \ep^2}{8} & \frac{2 \ep \left(6 \ep^2-1\right)}{45} & \frac{\pi
   \ep^2}{48}
\end{array}$
  \caption{The coefficients $w_{A,i}(\ep)$ and $w_{B,i}(\ep)$ appearing in Eqs.  \Ref{4.2.2.Ai} and \Ref{4.2.2.Bi}. }\label{4.2.w}
\end{table}

Now, using the expansions \Ref{4.2.2.Ai} and \Ref{4.2.2.Bi}, we can calculate from \Ref{4.2.2.Fic} the expansion
\be\label{4.2.2.Fi}\Phi_{(k,b)}(\om)=\sum_{i=0}^6 \Phi_{(k,b),i}\, \om^i+O(\om^6),
\ee
where
\be\label{4.2.2.Fk}\Phi_{(k,b),i}=w_{A,i}(\ep)+w_{B,i}(\ep).
\ee
For the four cases defined in Table \ref{4.2.cases}, these expansions read
\begin{eqnarray}\label{4.2.2.Fii}
\Phi_{(1,b)}&=&  \frac{\pi}{4} (   \ep -1) \omega ^2-\frac{1}{3} (\ep -1)^{3/2} \omega ^3+\frac{1}{36} (\ep -1)^{5/2} \omega ^5+O\left(\omega ^7\right),\\ \nn
\Phi_{(2,b)}&=&  \frac{1}{6} (\ep -1)^{3/2} \omega ^3-\frac{\pi}{32}  (\ep -1)^2  \omega ^4+\frac{1}{32} (\ep -1)^{5/2} \omega ^5+\frac{\pi}{768}    (\ep -1)^3 \omega ^6+O\left(\omega ^7\right),
\\ \nn
\Phi_{(3,b)}&=&  -\frac{\pi}{2}\frac{  (\ep -1) \ep}{ (\ep +1)} \omega ^2+\frac{2}{3} (\ep -1)^{3/2} \ep  \omega ^3-\frac{\pi}{8}    (\ep -1)^2 \ep ^2  \omega ^4
    \\\nn&&
    +\frac{2}{45} (\ep -1)^{5/2} \ep  \left(6 \ep ^2-1\right) \omega ^5+\frac{\pi}{48} (\ep -1)^3 \ep ^2 \omega ^6+O\left(\omega ^7\right),
\\ \nn
\Phi_{(4,b)}&=&  -\frac{\pi}{2}\frac{   (\ep -1)^2}{  (\ep +1)}\,\omega ^2+\frac{(\ep -1)^{3/2} \left(2 \ep ^2-1\right)}{6 \ep }\,\omega ^3-\frac{\pi}{32}   (\ep -1)^2 \ep ^2  \omega ^4
 \\\nn&&
 +\frac{(\ep -1)^{5/2} \left(48 \ep ^6-8 \ep ^4+2 \ep ^2+3\right) }{1440 \ep ^3}\,\omega ^5+\frac{\pi}{768}    (\ep -1)^3 \ep ^2 \omega ^6+O\left(\omega ^7\right).
\end{eqnarray}
In this way, the calculation of the contributions from region 2 to the low tempe\-r\-ature expansion is finished.
\subsubsection{The low frequency expansion}
Here we collect the contributions from the two regions calculated above. According to Eq. \Ref{4.2.12} one has to add them. It is seen that from region \a, Eq. \Ref{4.2.1.Fii},   only odd powers of $\om$ come, whereas from region \b, Eq. \Ref{4.2.2.Fii},   even powers come in too. Together, the expansions coefficients for the four cases in Table \Ref{4.2.cases} are
\bea\label{4.2.3.results}
\Phi_{(1)} &=&\frac{ \pi }{4}  (\ep -1) \om^2-\frac{1}{3} \left( \ep ^{3/2}-1\right) \om^3+\frac{1}{36} (\ep -1) \ep ^{3/2} \om^5+O\left(\om^7\right),\\
\Phi_{(2)} &=&\frac{1}{6} \left(\ep ^{3/2}-3 \sqrt{\ep }+2\right) \om^3-\frac{\pi }{32}   (\ep -1)^2 \om^4+\frac{(\ep -1)^2 (3 \ep -1) }{96 \sqrt{\ep }}\,\om^5
\nn\\&&
+\frac{\pi }{768} (\ep -1)^3 \om^6+O\left(\om^7\right),\nn\\
\Phi_{(3)} &=&-\frac{ \pi  (\ep -1) \ep  }{2 (\ep +1)}\,\om^2+\frac{1}{3} \left(2 \ep ^{5/2}-3 \ep ^{3/2}+1\right) \om^3-\frac{\pi }{8} (\ep -1)^2 \ep ^2 \om^4
\nn\\&&
+\frac{(\ep -1)\ep ^{3/2}}{180} (2 \ep  (12 \ep  (2 \ep -3)+5)+15) \om^5+\frac{\pi }{48} (\ep -1)^3 \ep ^2 \om^6+O\left(\om^7\right),\nn\\
\Phi_{(4)} &=&-\frac{ \pi  (\ep -1)^2 }{2 (\ep +1)}\,\om^2+\frac{1}{6} \left(\sqrt{\ep } (\ep  (2 \ep -3)-1)+2\right) \om^3-\frac{\pi }{32} (\ep -1)^2 \ep ^2 \om^4
\nn\\&&
+\frac{(\ep -1)^2 (2 \ep -1) \left(24 \ep ^3-7 \ep -3\right) }{1440 \sqrt{\ep }}\,\om^5+\frac{\pi }{768} (\ep -1)^3 \ep ^2 \om^6+O\left(\om^7\right). \nn
\eea
For the \elm case we have to add the polarizations. In case of  two dielectric half spaces we have to add cases 2 and 4 from \Ref{4.2.3.results},
\bea\label{4.2.3.ee}\Phi_{\ep,\ep}(\om)&\equiv&\Phi_{2}(\om)+\Phi_{4}(\om),
\nn\\               &=&
-\frac{ \pi  (\ep -1)^2 }{2 (\ep +1)}\,\om^2
+\frac{1}{3}
\left(\sqrt{\ep}-1 \right)\left(\ep^2+\ep^{3/2}-2\right)
   \om^3
\nn\\&&   -\frac{\pi }{32}  (\ep -1)^2 \left(\ep
   ^2+1\right) \om^4
   +\frac{(\ep -1)^2 (\ep +1) (\ep  (12
   \ep  (2 \ep -3)+29)-6)}{720 \sqrt{\ep }}\,\om^5
\nn\\&&   +\frac{ \pi }{768}
  (\ep -1)^3 \left(\ep ^2+1\right)
   \om^6+O\left(\om^7\right).
\eea
For one interface ideal conducting in front of a dielectric half space we have to add cases 1 and 3,
\bea\label{4.2.3.ie}
\Phi_{{\rm id.cond.},\ep}(\om)&\equiv&\Phi_{(1)}(\om)+\Phi_{(3)}(\om),
\\               &=&
-\frac{ \pi  (\ep -1)^2 }{4 (\ep +1)}\, \om^2+\frac{2}{3} \left(\ep ^{5/2}-2 \ep ^{3/2}+1\right) \om^3
-\frac{\pi }{8} (\ep -1)^2 \ep ^2 \om^4
\nn\\&& +\frac{(\ep -1) \ep ^{3/2}}{90} (\ep  (12 \ep  (2 \ep -3)+5)+10) \om^5
+\frac{ \pi }{48} (\ep -1)^3 \ep ^2 \om^6+O\left(\om^7\right).\nn
\eea
Inserted into Eq. \Ref{3.F6a}  these coefficients give the expansion of the free energy up to $T^7$, see Sect. 5.
As already mentioned, the limit $\ep\to\infty$, does not turn these coefficients into that if of the ideal conductor. Also, the two cases \Ref{4.2.3.ie} and \Ref{4.2.3.ee} are independent from one another.

\subsection{The plasma model}
The plasma model is described by the reflection coefficients \Ref{2.2.rq} or, equivalently, \Ref{2.2.reta} and by the permittivity \Ref{2.3.ep},
\be\label{4.3.ep}\ep^{\rm pl}(\om)=1-\frac{\Om^2}{\om^2}.
\ee
Since in the limit of infinite plasma frequency, $\Om\to\infty$, the free energy turns into that of the ideal conducting interfaces, there is no need here to introduce the cases used in the preceding subsection. Instead, we use on each interface its own plasma frequency, denoted by $\Oma$ and $\Omb$. The case of one ideal conducting interface can be restored afterwards by sending one of these  frequencies to infinity.

For the calculations, we divide the contributions according to regions \a\, and \b, defined in Sect. 3, Eq. \Ref{3.reg}.
\subsubsection{Region \a}
In this region we have a real $q$ and we use representation \Ref{3.fq},
\be\label{4.3.1.fi0}\phi_\a(i\om)=\int_0^\om dq\,q\, \ln\left(1-r_1r_2e^{-iq}\right),
\ee
where we have from \Ref{2.disp1} with the permittivity \Ref{2.3.ep} an imaginary $k_3=i\kappa$ with a real $\kappa=\sqrt{\Om^2-q^2}$. The reflection coefficients are given by Eqs. \Ref{2.2.rq}, where one has to insert for the corresponding interface, which we indicate by an additional index $i$ (i=1,2),
\be\label{4.3.1.rq}{\rE}_i=\frac{q-i\kappa_i}{q+i\kappa_i},\quad
            {\rM}_i=\frac{\left(1-\frac{\om^2}{\Omi^2}\right) q-i\kappa_i}
                         {\left(1-\frac{\om^2}{\Omi^2}\right) q+i\kappa_i},
\ee
with $\kappa_i=\sqrt{\Omi^2-q^2}$. For sufficiently small $\om$, $\om<\Omi$, these reflection coefficients are pure phase factors, i.e., their modula are equal to unity. Thus we can rewrite them in the form
\be\label{4.3.1.r}  {\rE}_i=e^{-i(\pi+2\alpha_{{\rm E}\,i})},\qquad
                    {\rM}_i=e^{-i 2\alpha_{{\rm E}\,i}  },
\ee
with
\be\label{4.3.}\alpha_{{\rm E}\,i}=\arctan\frac{q}{\sqrt{\Omi^2-q^2}},\quad
        \alpha_{{\rm M}\,i}=\arctan\frac{\om^2\sqrt{\Omi^2-q^2}}{(\Omi^2-\om^2)q^2}.
\ee
This allows to cast \Ref{4.3.1.fi0} in the form
\be\label{4.3.1.fi1}\phi_\a(i\om)=\int_0^\om dq\,q\, \ln\left(1- e^{-i\psi}\right),
\ee
where one has to insert
\be\label{4.3.1.psi} \begin{array}{rcll}
\psi_{\rm E}&=&2\alpha_{{\rm E}\,1}+2\alpha_{{\rm E}\,2}+q &\mbox{~~~(TE polarization)},
\\[8pt]
        \psi_{\rm M}&=&2\alpha_{{\rm M}\,1}+2\alpha_{{\rm M}\,2}+q &\mbox{~~~(TM polarization)}.
\end{array}\ee
Like in the case of ideal conducting interfaces, Eq. \Ref{4.1.ln}, this simple structure allows to rewrite the logarithm as
\be\label{4.3.1}\ln\left(1- e^{-i\psi}\right)=-i\frac{\psi}{2}+i\frac{\pi}{2}+\ln\left(2\sin\frac{\psi}{2}\right).
\ee
As before, the remaining log does not contribute since its argument does not change sign for sufficiently small $\om$ and we get with \Ref{3.Fi}
\be\label{4.3.1.FiE}\Phi_{\a,\, \rm TE}(\om)=\int_0^\om dq\,q \,(\psi_{\rm E}-\pi)
\ee
and
\be\label{4.3.1.FiM}\Phi_{\a,\, \rm TM}(\om)=\int_0^\om dq\,q \,(\psi_{\rm M}-\pi).
\ee
In these expressions, the integrations can be carried out explicitly and we come to
\bea\label{4.3.1.Fi2}
\Phi_{\a,\, \rm TE}(\om)&=&-\frac{\pi}{2}\,\om^2+\frac{1}{3}\,\om^3+
2\Oma^2h_{\rm E}\left(\frac{\om}{\Oma}\right)
+2\Omb^2h_{\rm E}\left(\frac{\om}{\Omb}\right),
\nn\\[4pt]
\Phi_{\a,\, \rm TM}(\om)&=&-\frac{\pi}{2}\,\om^2+\frac{1}{3}\,\om^3+
2\Oma^2h_{\rm M}\left(\frac{\om}{\Oma}\right)
+2\Omb^2h_{\rm M}\left(\frac{\om}{\Omb}\right),
\eea
where
\bea\label{4.3.1.hE}h_{\rm E}(z)&=&\int_0^z dq\,q\,\arctan\frac{q}{\sqrt{z^2-q^2}}
\nn\\&=&\frac{z\sqrt{1-z^2}}{4}+\frac{2z^2-1}{4}\arctan z
\eea
and
\bea\label{4.3.1.hM}h_{\rm M}(z)&=&\int_0^z dq\,q\,\arctan
\frac{z^2\sqrt{z^2-q^2}}{q(1-z^2)}
\nn\\&=&\frac{\pi}{4}\frac{z^4}{1-2z^2}+\frac{z^2(z^2-1)}{1-2z^2}\arcsin z
\eea
result from the integrations in \Ref{4.3.1.FiE} and \Ref{4.3.1.FiM}.

The expansion of these functions,
\bea\label{4.3.1.hX}h_{\rm E}(z)&=&\sum_{i\ge3}h_{{\rm E},\,3}\,z^i=
    \frac{1}{3}\,z^3+\frac{1}{30}\,z^5+\frac{3}{280}\,z^7+O(z^9),
\\ \nn      h_{\rm M}(\,z)&=&\sum_{i\ge3}h_{{\rm M},\,3}\,z^i=
    z^3-\frac{\pi}{4}\,z^4+\frac{7}{6}\,z^5
            -\frac{\pi}{2}\,z^6+\frac{269}{120}\,z^7+O(z^8),
\eea
can be used in \Ref{4.3.1.Fi2} to obtain the expansions
\be\label{4.3.1.Fia}\Phi_{\a,\,{\rm X}}(\om)=
    -\pi\om^2+\left(\frac13+\frac12\left(\frac{1}{\Oma}+
    \frac{1}{\Omb}\right)h_{{\rm X},\,3}\right)\,\om^3+
    \sum_{i\ge4}\left(\frac{1}{\Oma^{i-2}}+ \frac{1}{\Omb^{i-2}}\right)h_{{\rm X},\,i}\,\om^i,
\ee
where we used the notation X for TE and TM not to write down nearly the same formula twice. From this representation it is seen  that the ideal conductor result \Ref{4.1.Fi} can be reobtained in the limit of both plasma frequencies becoming large.

\subsubsection{Region \b}
In this region we have a real $\eta$ and, from \Ref{2.2.kappa} with \Ref{4.3.ep} inserted, and a real $\kappa=\sqrt{\Om^2+\eta^2}$.
The reflection coefficients \Ref{2.2.reta} are, explicitly written,
\be\label{4.3.2.r} \rE=\frac{\eta-\sqrt{\Om^2+\eta^2}}{\eta+\sqrt{\Om^2+\eta^2}},\qquad
\rM=\frac{\left(1-\frac{\Om^2}{\om^2}\right)\eta-\sqrt{\Om^2+\eta^2}}
{\left(1-\frac{\Om^2}{\om^2}\right)\eta+\sqrt{\Om^2+\eta^2}}.
\ee
In the case, the two interfaces have different plasma frequencies, we have to insert $\Oma$ and $\Omb$ into the corresponding
%
reflection coefficients $r_i$ $(i=1,2)$, Eq. \Ref{2.2.reta} and in
\be\label{4.3.2.ka}\kappa_i=\sqrt{\Omi^2+\eta^2}\qquad (i=1,2).
\ee
These coefficients are real. Thus the function \Ref{3.f22},
\be\label{4.3.2.fi}\phi_\b(i\om)=
        \int_0^\infty\,d\eta\,\eta\,\ln\left(1-r_1r_2\,e^{-\eta}\right),
\ee
seems to be real too. If this would be the case, it would not contribute to
\be\label{4.3.2.Fi}\Phi_\b(\om)=i(\phi_\b(i\om)-\cc).
\ee
However, it happens that the argument of the logarithm in \Ref{4.3.2.Fi} changes sign not only for a finite $\om$ as in the case of an ideal conductor, but also for arbitrarily small $\om$.

Physically this can be understood from the spectrum. We consider real $\om$ with imaginary both momenta, $q=i\eta$ inside and $k_3=i\kappa$ outside the gap. The plasma model is known to have, for the TM polarization, such excitations, the surface modes (or surface plasmons). Further, the argument of the logarithm in \Ref{4.3.2.fi} is proportional to the transmission coefficient of the system of two interfaces, thus it has zeros at the wave numbers corresponding to these excitations at a given frequency $\om$. Moreover,   the argument of the logarithm has poles corresponding to the surface plasmons on the interfaces taken individually, where their reflection coefficients $r_i$ (i=1,2) become infinite.

Since there are no surface plasmons for the TE polarization, we can be sure that the corresponding logarithm in \Ref{4.3.2.fi} stays real and that this polarization does not contribute to \Ref{4.3.2.Fi}.
For the TM polarization we rewrite \Ref{4.3.2.fi} using \Ref{2.2.reta},
%
\bea\label{4.3.2.fi1}  \phi_\b(i\om)&=&\int_0^\infty\,d\eta\,\eta\,\left[
    -\ln\left(\ep_1\eta+\kappa_1\right)-\ln\left(\ep_2\eta+\kappa_2\right)
 \right.\\\nn&&\left.   +\ln\left(
        \left(\ep_1\eta+\kappa_1\right)\left(\ep_2\eta+\kappa_2\right)
        -\left(\ep_1\eta-\kappa_1\right)\left(\ep_2\eta-\kappa_2\right)\,e^{-\eta}
                    \right)
\right].
\eea
Here, the arguments of the logarithms have no poles but only zeros. We denote the wave number $\eta$ for the single plasmons by $\eta_{{\rm single}, i}$ $(i=1,2)$. These follow from the poles of $r_i$, i.e. these are solutions of
\be\label{4.3.2.}\ep_i \eta+\kappa_i=0 \qquad (i=1,2).
\ee
Using \Ref{4.3.ep} and \Ref{4.3.2.ka}, these equations can be solved for $\eta$ resulting in
\be\label{4.3.2.etasi}\eta_{{\rm single},i}=\frac{\om^2}{\sqrt{\Omi^2-2\om^2}}
\qquad (i=1,2).
\ee
The wave numbers of the plasmons in the system of two interfaces we denote by $\eta_{{\rm symm.}}$ and $\eta_{{\rm antisymm.}}$. These notations account for the properties of the corresponding wave functions to  have a symmetry in case of equal plasma frequencies.
These wave numbers are solutions of the equation
\be\label{4.3.2.eq}1-r_1r_2\,e^{-\eta}=0,
\ee
or, equivalently, of equating the argument of the third logarithm in \Ref{4.3.2.fi1} to zero.

It is known (see, for example Fig. 2 in \cite{bord12-85-025005}) that $\eta_{{\rm antisymm.}}$ is not real for small $\om$. Therefore it does not contribute in region \b.
The solution $\eta_{{\rm symm.}}$ of Eq. \Ref{4.3.2.eq} exists for arbitrarily small $\om$ and it has an expansion
\be\label{4.3.2.exp}\eta_{{\rm symm.}}=\sum_{j\ge0}a_j\,\om^j,
\ee
whose coefficients can be easily calculated by inserting \Ref{4.3.2.exp} into \Ref{4.3.2.eq} and expanding in powers of $\om$. One obtains
\bea\label{4.3.2.etasy}\eta_{{\rm symm.}}&=&
    \sqrt{2}\left(\frac{1}{\Oma}+\frac{1}{\Omb}\right) \om
\nn\\&&    +
\frac{1}{6\sqrt{2}}\left(\frac{6}{\Oma^4}+\frac{6}{\Oma^3
   \Omb}+\frac{6}{\Oma^3}+\frac{1}{\Oma^2}+\frac{6}{\Oma
   \Omb^3}-\frac{1}{\Oma
   \Omb}
   \right.\nn\\&&\left.+\frac{6}{\Omb^4}+\frac{6}{\Omb^3}+\frac{1}{\Omb
   ^2}\right) \sqrt{\frac{\Oma \Omb}{\Oma+\Omb}}
                   \, \, \om^3+O(\om^5).
\eea
In case of equal frequencies, this expression simplifies,
\bea\label{4.3.2.syeq}{\eta_{{\rm symm.}}}_{{ |}_{\Oma=\Omb=\Om}}&=&
2 \frac{\om}{\sqrt{\Om}}
+\left(\frac{2}{\Om^2}+\frac{1}{\Om}+\frac{1}{12}\right)
   \left(\frac{\om}{\sqrt{\Om}}\right) ^3
\\\nn&&+\left(\frac{1}{\Om^4}+\frac{3}{\Om^3}+\frac{7}{6
   \Om^2}+\frac{1}{8 \Om}+\frac{11}{2880}\right) \left(\frac{\om}{\sqrt{\Om}}\right)
   ^5+O\left(\om ^7\right).
\eea
The analytic continuation in $\phi_\b(i\om)$, Eq. \Ref{4.3.2.fi},  is to be taken starting from large $\eta$ where the expression is real because of the exponential $\exp(-\eta)$. When the argument of a logarithm, for decreasing $\eta$,  passes zero, the logarithm acquires, from each passing, an addendum of $i\pi$. Thus we have in \Ref{4.3.2.Fi}
\bea\label{4.3.2.Fi1}\Phi_\b(\om)&=&-2\pi\left[
        -\int_0^{\eta_{{\rm single}, 1}}d\eta\,\eta
        -\int_0^{\eta_{{\rm single}, 2}}d\eta\,\eta
        +\int_0^{\eta_{{\rm symm.}}}d\eta\,\eta      \right]
,\nn\\[8pt]&=&    -\pi\left(
            \eta_{{\rm symm.}}^2-\eta_{{\rm single}, 1}^2-\eta_{{\rm single}, 1}^2
                    \right).
\eea
Inserting \Ref{4.3.2.etasi} and \Ref{4.3.2.etasy}, after re-expansion, we get
\bea\label{4.3.2.Fib}
\Phi_\b(\om)&=&
-2\pi\left(\frac{1}{\Oma}+\frac{1}{\Omb}\right)\,\om^2+
\frac{\pi}{3}\left(
2 \left(\frac{1}{\Oma^2}+\frac{1}{\Omb^2}\right)+\frac{1}{\Oma\Omb}
\right.\nn\\\nn&&\left.
-6
   \left(\frac{1}{\Oma^4}+\frac{1}{\Oma^3
   \Omb}+\frac{1}{\Oma^3}+\frac{1}{\Oma
   \Omb^3}+\frac{1}{\Omb^4}+\frac{1}{\Omb^3}\right)
   \right)\,\om^4
+O(\om^6),
\eea
which is, for equal plasma frequencies,
\be\label{4.3.2.Fibeq}{\Phi_\b(\om)}_{{\Oma |}_{\Oma=\Omb=\Om}}=
-4\pi\frac{\om^2}{\Om}+\frac{5\pi}{3}
    \left(1-\frac{12}{5}\frac{1}{\Om}-\frac{24}{5}\frac{1}{\Om^2}\right)
                \frac{\om^4}{\Om^2}+O(\om^6).
\ee
It is seen that, in the limit of large plasma frequencies, there is no contribution from region \b.

\subsubsection{The low frequency expansion}
Adding the contributions from both regions we get the low temperature expansion for the plasma model. For the TE polarization, only region \a\, contributes to $\Phi(\om)$ and we have only to rewrite \Ref{4.3.1.Fia} adding the index 'TE',
\be\label{4.3.3.FiTE}\Phi_{\rm TE}(\om)=-\frac{\pi}{2}\,\om^2
+\left(\frac13+\frac23\left(\frac{1}{\Oma}+\frac{1}{\Omb}\right)\right)\om^3
+\frac{1}{15}\left(\frac{1}{\Oma^3}+\frac{1}{\Omb^3}\right)\om^5
+O(\om^7).
\ee
For the TM polarization we have to add \Ref{4.3.1.Fia} and \Ref{4.3.2.Fib},
\bea\label{4.3.3.FiTM}\Phi_{\rm TM}(\om)&=&
-\left(\frac{\pi}{2}+2\left(\frac{1}{\Oma}+\frac{1}{\Omb}\right)\right)\,\om^2
+\left(\frac{1}{3}+2\left(\frac{1}{\Oma}+\frac{1}{\Omb}\right)\right)\,\om^3
\nn\\&&
+\frac{\pi}{3 }
        \left(1-\frac{6 \Omb}{\Oma^3}
        -\frac{6   \Omb}{\Oma^2}
        -\frac{6}{\Oma^2}-\frac{6   \Oma}{\Omb^3}
        -\frac{6 \Oma}{\Omb^2}
        +\frac{\Oma}{2   \Omb}
\right.\nn\\&&\left.
           +\frac{\Omb}{2 \Oma}
        -\frac{6}{\Omb^2}
   \right)\frac{\om^4}{\Oma\Omb}
+\frac{7}{3}\left(\frac{1}{\Oma^3}+\frac{1}{\Omb^3}\right)\om^5
+O(\om^7),
\eea
which also simplifies,
\bea\label{4.3.3.FiTMeq}{\Phi_{\rm TM}(\om)}_{\Oma=\Omb=\Om}&=&
-\frac{\pi}{2}\left(1+\frac{8}{\Om}\right) {\om^2}
+\frac{1}{3}\left(1+\frac{12}{\Om}\right) {\om^3}
\nn\\&&+\frac{2\pi}{3}
    \left(1- \frac{6}{\Om}- \frac{12}{\Om^2}\right)
                \frac{\om^4}{\Om^2}+\frac{14}{3}\frac{\om^5}{\Om^3}+O(\om^6),
\eea
in case of equal plasma frequencies.

Finally, for the complete \elm case, we have to add the polarizations, Eqs. \Ref{4.3.3.FiTE} and \Ref{4.3.3.FiTM}, and get
\bea\label{4.3.3.Fi}
{\Phi_{\rm ED} (\om)}&=&
- {\pi}\left(1 +2\left(\frac{1}{\Oma}+\frac{1}{\Omb}\right)\right)\,\om^2
+\left(\frac{2}{3}+\frac{8}{3}\left(\frac{1}{\Oma}+\frac{1}{\Omb}\right)\right)\,\om^3
\nn\\&&
+\frac{\pi}{3 }        \left(
    1-\frac{6 \Omb}{\Oma^3}-\frac{6
   \Omb}{\Oma^2}-\frac{6}{\Oma^2}-\frac{6
   \Oma}{\Omb^3}-\frac{6 \Oma}{\Omb^2}+\frac{\Oma}{2
   \Omb}
\right.\nn\\&&\left.+\frac{\Omb}{2 \Oma}-\frac{6}{\Omb^2}
           \right)\frac{\om^4}{\Oma\Omb}
+\frac{12}{5}\left(\frac{1}{\Oma^3}+\frac{1}{\Omb^3}\right)\om^5
+O(\om^6).
\eea
The coefficients of this expansion, identified according to Eq. \Ref{3.Fi2}, and being inserted into Eq. \Ref{3.F6a}, give the low temperature expansion for the plasma model.

As special case we note the corresponding formula,
\bea\label{4.3.3.Fieq}{\Phi_{\rm ED} (\om)}_{\Oma=\Omb=\Om}&=&
- {\pi} \left(1+\frac{4}{\Om}\right) {\om^2}
+\frac{2}{3}\left(1+\frac{8}{\Om}\right) {\om^3}
\nn\\&&+\frac{2\pi}{3}
    \left(1- \frac{6}{\Om}- \frac{12}{\Om^2}\right)
                \frac{\om^4}{\Om^2}
                +\frac{24}{5}\frac{\om^5}{\Om^3}+O(\om^6),
\eea
for equal plasma frequencies and
\bea\label{4.3.3.Fiid}{\Phi_{\rm ED} (\om)}_{\Oma=\infty,\,\Omb=\Om}&=&
- {\pi} \left(1+\frac{2}{\Om}\right) {\om^2}
+\frac{2}{3}\left(1+\frac{4}{\Om}\right) {\om^3}
\nn\\&&+\frac{\pi}{6}
    \left(1- \frac{12}{\Om}- \frac{12}{\Om^2}\right)
                \frac{\om^4}{\Om^2}
                +\frac{12}{5}\frac{\om^5}{\Om^3}+O(\om^6),
\eea
for one interface ideally conducting.

\subsection{The Drude model}
The reflection coefficients for the Drude model are given by Eq. \Ref{2.2.rq} or by \Ref{2.2.reta} with the permittivity given by Eq. \Ref{2.4.ep}.
As already mentioned in the Introduction, the limit $\ga\to0$  of vanishing dissipation parameter does not reproduce the free energy of the plasma model and there is a problem with thermodynamics. In the following subsections we consider first the case $T\to0$ for fixed $\ga$ and, afterwards, the case of vanishing $\ga$.

\subsubsection{Fixed dissipation parameter}
In this subsection we consider a fixed dissipation parameter $\ga$. In that case the low temperature expansion of the free energy is given by Eq. \Ref{3.F6a} with coefficients following from the expansion    \Ref{3.Fi2}.

As already mentioned,   the limit $\ga\to0$ does not reproduce the free energy of the plasma model.
Thus, even for fixed dissipation parameter,  the low temperature expansion of the free energy in the Drude model is quite different from that of the plasma model and the ideal conductor results, on one or on both interfaces, and it  cannot be obtained as some limiting case. For this reason, we consider the four cases introduced in Sect. 4.2. in Table \Ref{4.2.cases}. Also we restrict ourselves to the order $\om$, i.e., to the order $T^2$,  which is the leading order for this model. On the one side, higher orders are technically elaborate, on the other side, these are hardly of any use.

For the calculations we use the scheme of two regions introduced in Sect. 3, Eq. \Ref{3.reg}.
In region \a\, we use representation \Ref{3.fq} with $q$ as integration variable,
\be\label{4.4.1.fi}\phi_\a (i\om)=\int_0^\om dq\,q\,\ln\left(1-r_1r_2e^{-iq}\right).
\ee
For the reflection coefficients one has to insert from \Ref{2.2.rq} according to the case from Table \Ref{4.2.cases} with $k_3=\sqrt{(\ep^{\rm Dr}(-\om)-1)\om^2+q^2}$ and the permittivity 
\be\label{4.4.1.ep}\ep^{\rm Dr}(-\om)=1-\frac{\Om^2}{\om(\om-i\ga)}
\ee
in place of \Ref{2.4.ep} accounting for \Ref{3.AW}.

Now we make in Eq. \Ref{4.4.1.fi} the substitution $q\to\om q$. We get a factor $\om^2$ in front. The remaining integration remains finite in the limit $\om\to0$. Thus, the contribution  from region \a\, starts from $\om^2$ and is beyond we are interested in.

In region \b\, we have a real $\eta$ and from Eq. \Ref{2.2.kappa} and \Ref{4.4.1.ep},
\be\label{4.4.1.ka}\kappa=\sqrt{\Om^2 \frac{\om}{\om-i \ga}+\eta^2},
\ee
which is complex for all $\eta$. Thus, we have in \Ref{3.f22}
\be \label{4.4.1.fib}\phi_\b (i\om)=\int_0^\infty
    d\eta\,\eta\,\ln\left(1-r_1r_2e^{-\eta}\right),
\ee
where we used the convention introduced with Eq. \Ref{4.2.ki}. Here, in distinction from the previous cases,   the whole integration region contributes. It turns out that it is not possible to obtain an expansion, for $\om\to0$, by simple substitution and expansion of the integrand.

First we consider case 1 as defined in Table \Ref{4.2.cases} and in Eq. \Ref{4.2.ki}. The function \Ref{4.4.1.fib} takes the specific form
\be \label{4.4.1.fi1}\phi_{(1,b)} (i\om)=\int_0^\infty
    d\eta\,\eta\,\ln\left(1+\rE e^{-\eta}\right)
\ee
with
\be\label{4.4.1.rE} \rE=\frac{\eta-\sqrt{\Om^2\frac{\om}{\om-i \ga}+\eta^2}}
                                {\eta+\sqrt{\Om^2\frac{\om}{\om-i \ga}+\eta^2}}.
\ee
For the expansion we use Appendix A. The integral in \Ref{4.4.1.fi1} is the same as in \Ref{A.1} with the substitutions
\be\label{4.4.1.A}r_0\to -1,\qquad \Omega^2\to \Om^2\frac{\om}{\om-i \ga}.
\ee
Its expansion for small $\Omega$ is given by Eq. \Ref{A.2} with the coefficients \Ref{A.r-1}. From the leading order term of the expansion \Ref{A.2}  we get for \Ref{4.4.1.fi1} in leading order in $\om$
\be\label{4.4.1.fi4}\phi_{(1,b)} (i\om)=\frac{\Om^2}{4i\ga}
        \left( \gamma_{\rm E}-1+\frac12\ln\frac{4\Om^2 \om}{i\ga}\right)\om
        +\frac{\Om^3\, \om^{3/2}}{3(-i\ga)^{3/2}}
        +O(\om^2).
\ee
After inserting into \Ref{3.Fi}  we get finally
\be\label{4.4.1.Fi}\Phi_{(1,b)}(\om)=\frac{\Om^2}{2\ga}
    \left( \gamma_{\rm E}-1+\frac12\ln\frac{4\Om^2 \om}{ \ga}\right)\,\om
    - \frac{\Om^3\,\om^{3/2}}{3\sqrt{2}\,\ga^{3/2}}
+O(\om^{2}).
\ee
Since there is no contribution from region \a\, to this order, the above is the complete result in leading order. This is just the case involving a logarithmic contribution shown in the expansions \Ref{3.Fi2} and \Ref{3.F6a}. Comparing these with \Ref{4.4.1.Fi}, we can identify the coefficients,
\bea\label{4.4.1.Fit}\Phi_{(1),1}&=&  \left(\gamma_{\rm E}-1+\frac12\ln\frac{4\Om^2}{\ga}\right)
                    \frac{\Om^2}{4\ga},
\nn\\   \tilde{\Phi}_{(1),1}&=& \frac{\Om^2}{4\ga},\quad \Phi_{(1),\frac{3}{2}}=-\frac{\Om^3}{3\sqrt{2}\,\ga^{3/2}},
\eea
which, after insertion into \Ref{3.F6a}, give the leading order terms in the low temperature expansion proportional to $T^2$, $T^2\ln T$ and $T^{3/2}$.

Next we consider case 2 from Table \ref{4.2.cases}. Here we have
\be \label{4.4.1.fi2}\phi_{(2,b)} (i\om)=\int_0^\infty
    d\eta\,\eta\,\ln\left(1-\rE^2 e^{-\eta}\right),
\ee
with $\rE$ given by Eq. \Ref{4.4.1.rE}. In this case, for the leading order, it is possible to make the substitution $\eta\to \Om\sqrt{\frac{\om}{\om+i \ga}}\,\eta$ in the integral. After that the integrand can be expanded,
\be\label{4.4.1.fi2a}\phi_{(2,b)} (i\om)=\frac{\Om^2\,\om}{\om+i \ga}\int_0^\infty
    d\eta\,\eta\,\ln\frac{4\eta\sqrt{\eta^2+i}}{(\eta+\sqrt{\eta^2+i})^2}+O(\om^{3/2}).
\ee
The remaining integration is simple and after inserting  into \Ref{3.Fi} the result is
\be\label{4.4.1.Fi2}\Phi_{(2)}(\om)=\frac{2\ln 2-1}{2}\frac{\Om^2}{\ga}\,\om+O(\om^{3/2}),
\ee
remembering that there is no contribution from region \a.

The corresponding contribution to the free energy (the first term in Eq. \Ref{4.4.1.FTE} below) was obtained in \cite{brev04-QFEXT}. In \cite{hoye07-75-051127}, see also \cite{brev08-41-164017,elli09-161-012010}, also the next order, which is proportional to $\om^{3/2}$, was calculated. This coefficient can be obtained exactly using the same method as above for the case 1. For this we rewrite Eq. \Ref{4.4.1.fi2} in the form
\be \label{4.4.1.ho1}\phi_{(2,b)} (i\om)=
\int_0^\infty  d\eta\,\eta\,\ln\left(1-\rE  e^{-\eta/2}\right)
+\int_0^\infty  d\eta\,\eta\,\ln\left(1+\rE  e^{-\eta/2}\right),
\ee
reducing the problem, in this way, to the previous case. In both integrals we make the substitution $\eta\to 2\eta$. We get a factor of 4 in front of the integral and have to substitute $\Omega\to\Omega/2$ when applying Eq. \Ref{A.1}. 
Now we use the expansion \Ref{A.2} from the Appendix A for the two integrals with $r_0=+1$ for the first and $r_0=-1$ for the second, together with the substitution of $\Omega$ according to Eq. \Ref{4.4.1.A}. We restrict ourselves to the next-to-leading order (for higher orders one would need to calculate the corresponding contributions from region \a\, too) and obtain from $a_2$ and $a_3$ in \Ref{A.r-1} and \Ref{A.r1},
\be\label{4.4.1.fi2b}
        \phi_{(2,b)}(i\om)=\frac{1-2\ln 2}{4}\frac{\Om^2\,\om}{-i\ga}
                +\frac{\Om^3}{12(-i\ga)^{3/2}}\,\om^{3/2}+O(\om^2),
\ee
resulting with \Ref{3.zush} in
\be\label{4.4.1.Fi2n}\Phi_{(2)}(\om)=\frac{2\ln 2-1}{2}\frac{\Om^2}{\ga}\,\om
        -\frac{1}{6\sqrt{2}}\frac{\Om^3}{\ga^{3/2}}\,\om^{3/2}+
        O(\om^{2}).
\ee
In both contributions we observe a minus sign appearing from  \Ref{3.Fi}.
The leading order coincides, of course, with the result from the integration in \Ref{4.4.1.fi2a}.
Using Eq. \Ref{3.zeta}, the corresponding contribution to the free energy,  resulting from the TE polarization, reads
\be\label{4.4.1.FTE} \Delta_T\F^{\rm TE}=
    \frac{2\ln2-1}{48}\frac{\Om^2}{\ga}\,T^2
    -\frac{\zeta_{\rm R}(5/2)}{16\sqrt{2}\pi^{3/2}}\frac{\Om^{3}}{\ga^{3/2}}\,a\,T^{5/2}+O(T^3),
\ee
where we restored the dependence on $a$ using \Ref{2.restore}-\Ref{2.restore4}. The numerical coefficient
$\zeta_{\rm R}(5/2)/(16\sqrt{2}\pi^{3/2})=-\sqrt{2\pi}\zeta_{\rm R}(-3/2)/6\sim 0.00191$ coincides with the one in \cite{hoye07-75-051127} up to the sign. We mention that, using the formulas from Appendix A, higher orders can be written down too. These formulas can be used also to obtain the higher orders in case 1,
\be\label{4.4.1.Fi1n}\Phi_{(1)}(\om)=
    \frac{\Om^2}{2\ga}
    \left( \gamma_{\rm E}-1+\frac12\ln\frac{4\Om^2 \om}{ \ga}\right)\,\om
    -\frac{1}{3\sqrt{2}}\frac{\Om^3}{\ga^{3/2}}\,\om^{3/2}+
    O(\om^{2}).
\ee
The remaining two cases from Table \Ref{4.2.cases}, which belong to the TM polarization, are easier if restricting to the leading order. Here we have to insert $\rM$, Eq. \Ref{2.2.reta}, with the permittivity \Ref{4.4.1.ep} into \Ref{3.f22}. The direct expansion of the integrand for $\om\to0$ is possible. Using, for $\om\to0$,
\be\label{4.4.1.ln}\ln\left(1-\rM e^{-\eta}\right)=
                        \frac{2i\ga}{(e^\eta-1)\Om^2}\,\om+\dots\,,
\ee
the integration over $\eta$ can be carried out and one comes to
\be\label{4.4.1.Fi3}\Phi_{(3)}(\om)=
            -\frac{2\pi^2\ga}{3\Om^2}\,\om+\dots\,.
\ee
The last, case 4, can be treated in complete analogy and we get
\be\label{4.4.1.Fi4}\Phi_{(4)}(\om)=
            -\frac{4\pi^2\ga}{3\Om^2}\,\om+\dots\,,
\ee
which completes the calculation of the low temperature expansion \Ref{3.F6a} for the Drude model with fixed dissipation parameter.

\subsubsection{Vanishing dissipation parameter}
As already mentioned in the Introduction, there is a problem with the Drude model for vanishing dissipation parameter $\ga$. The problem has two aspects. First, consider the entropy $S$, related by the usual relation \Ref{1.S} with the free energy \Ref{2.F}, for $T\to0$. In case, the relaxation parameter depends on temperature, $\ga(T)$, and it decreases, for $T\to0$, not slower than the first power of $T$,
\be\label{4.4.2.ga}\ga(T)\raisebox{-4pt}{${\sim \atop T\to0}$}T^\alpha\qquad(\alpha>1),
\ee
there is a linear term in the expansion for $T\to0$,
\be\label{4.4.2.FT}\F=E_0+\F_1T+\dots\,,
\ee
of the free energy $\F^{\rm Dr}$ of the Drude model which causes a non vanishing contribution to the entropy for $T\to0$,
\be\label{4.4.2.S0}\lim_{T\to0} S=-\F_1.
\ee
Such non zero $\F_1$ constitutes a violation of the third law of thermodynamics (Nernst' heat theorem). It must be mentioned that the above discussion applies to a dissipation parameter obeying \Ref{4.4.2.ga}. The importance of this case for the Casimir effect was discussed in \cite{BKMM}, Chapt. 14.3.2.

The other aspect appears if considering the free energy of the Drude model for vanishing relaxation parameter, $\ga\to0$, at fixed temperature $T$. While the permittivity \Ref{2.3.ep} turned into that of the plasma model,
\be\label{4.4.2.ep}\ep^{\rm Dr}\raisebox{-4pt}{${\sim \atop \ga\to0}$}\ep^{\rm pl},
\ee
and  so do the reflection coefficients \Ref{2.2.rq}, the free energy does not turn into the free energy $\F^{\rm pl}$ of the plasma model,
\be\label{4.4.2.F}\F^{\rm Dr}\raisebox{-4pt}{${\sim \atop \ga\to0}$}\F^{\rm pl}+\F_1 T,
\ee
keeping, in the limit $\ga\to0$, an additional contribution, where $\F_1$ is just the same as in Eq. \Ref{4.4.2.S0}. This contribution is, from a physical point of view, unsatisfactory since a small, or even vanishing, dissipation should have a correspondingly small effect on measurable quantities like the force ($\F_1$ depends on the separation $a$).

It is interesting to mention that the experimental verification of the Casimir force, in a number of experiments, favors the plasma model and rules out the Drude model \cite{klim09-81-1827}, while other experiments support the Drude model \cite{garc12-109-027202} (but have been questioned in \cite{bord12-109-199701}). In the  experiments the dissipation parameter is small, a typical value is  $\ga$= 0.035 eV while $\Om$ = 9 eV for gold. Now, if assuming that the correct formula, describing the Casimir force in case of dissipation, is perturbative in $\ga$, it would be clear that the dissipation gives only a small addendum wich does not show up in the experiments to date.

\begin{figure}
  \includegraphics[width=8 cm]{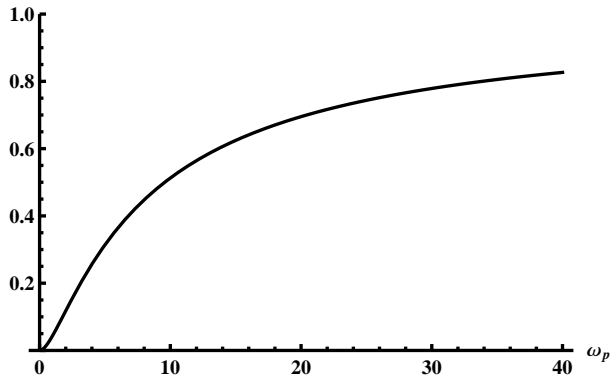}\\
  \caption{The function $\F_1(\Om)$ normalized to its value $\F_1(\infty)=-\zeta(3)/4\pi^2$, Eq. \Ref{4.4.2.exp},  at infinite $\Om$.}\label{4.4.fig}
\end{figure}

There are two ways to derive $\F_1$. The {\bf first way} uses the representation \Ref{2.F} in terms of Matsubara frequencies. It is a property of the reflection coefficients \Ref{2.2.reta},
\be\label{4.4.2.reta}\rE^{\rm Dr}=\frac{\eta-\kappa_l}{\eta+\kappa_l},\quad
                    \rM^{\rm Dr}=\frac{\ep^{\rm Dr}(i\xi_l) \eta-\kappa_l}{\ep^{\rm Dr}(i\xi_l) \eta+\kappa_l},
\ee
with, from Eq. \Ref{2.4.ep2},
\be\label{4.4.2.ep0}\ep^{\rm Dr}(i\xi_l)=1+\frac{\Om^2}{\xi_l(\xi_l+\ga)},
\ee
and
\be\label{4.4.2.kal}\kappa_l=\sqrt{\Om^2\frac{\xi_l}{\xi_l+\ga}+\eta^2},
\ee
to have a sufficiently different behavior for $l\ne0$ and $\l=0$. For $\l=0$ we have $\kappa_0=\eta$ and $\ep^{\rm Dr}(0)=\infty$. Hence, for the reflection coefficients
\be\label{4.4.2.r0}\rE^{\rm Dr}=0,\quad
                    \rM^{\rm Dr}=1\qquad (l=0)
\ee
holds, while in the plasma model, where we have still $\ep^{\rm pl}(0)=\infty$, but from \Ref{2.3.kappa},
\be\label{4.4.2.kal0}\kappa_l=\sqrt{\Om^2+\eta^2},
\ee
a   $\kappa_0\ne \eta$,
the relations
\be\label{4.4.2.r0pl}\rE^{\rm pl}=\frac{\eta-\sqrt{\Om^2+\eta^2}}
                                        {\eta+\sqrt{\Om^2+\eta^2}},\quad
                    \rM^{\rm pl}=1\qquad (l=0)
\ee
hold. In this way, the contribution from $\l=0$ to the Matsubara sum is different for both models. It must be mentioned, that in the Matsubara representation, the $(l=0)$-contribution to the TE polarization is the only place where  this difference shows up. All other contributions to the free energy in the Drude model turn into their counterparts in the plasma model.
Thus, for $\ga\to0$, the difference in the free energies between both models,
\be\label{4.4.2.F1}\lim_{\ga\to0}\left(\F^{\rm Dr}-\F^{\rm pl}\right)=\F_1 \,T,
\ee
results just from the $(l=0)$-contribution in the TE polarization to the free energy of the plasma model, see \cite{beze04-69-022119}. The contribution from the Drude model vanishes along with the reflection coefficient $\rE^{\rm Dr}$, Eq. \Ref{4.4.2.r0}. This difference can be calculated easily from Eq. \Ref{2.F},
\be\label{4.4.2.F2}\F_1=-\frac{1}{4\pi^2}\,\phi(0)
\ee
with, from \Ref{2.fi},
\be\label{4.4.2.fi0}\phi(0)=\int_0^\infty dk\,k\,\ln\left(1-\r(\Om,k)^2 \,e^{-k}\right).
\ee
Here we used, for $l=0$, \Ref{4.4.2.kal}  and $\eta=k$ which follows from \Ref{2.disp3} and \Ref{2.2.reta}.
Eqs. \Ref{4.4.2.F2} and \Ref{4.4.2.fi0} were first obtained in \cite{beze04-69-022119}.
For the reflection coefficient in \Ref{4.4.2.fi0} we introduced with
\be\label{4.4.2.Om}\r(\Om,k)=\frac{k-\sqrt{\Om^2+k^2}}
                                {k+\sqrt{\Om^2+k^2}},
\ee
a separate notation, which will be used several more times below. It results from the TE polarization, Eq. \Ref{4.3.2.r},
\be\label{4.4.2.Om1} \r(\Om,\eta)=\rE^{\rm pl}.
\ee
The function $\F_1$ depends on one variable $\Om$ only. We make this fact explicit by writing
\be\label{4.4.2.F3}\F_1\equiv\F_1(\Om)=-\frac{1}{4\pi^2}\int_0^\infty dk\,k\,\ln\left(1-\r(\Om,k)^2 \,e^{-k}\right).
\ee
This function can be easily calculated numerically since the integral is rapidly converging. A plot is shown in Fig. \ref{4.4.fig}.
The asymptotic of this function for large argument, which by means of \Ref{2.3.skin} corresponds to small skin depths, can be easily obtained from expanding the integrand in \Ref{4.4.2.F3} with subsequent integration,
\be\label{4.4.2.exp}\F_1=
\frac{-1}{4\pi^2}\left(
 \zeta_{\rm R}(3)
-8\zeta_{\rm R}(3)\frac{1}{\Om}
+48\zeta_{\rm R}(3)\frac{1}{\Om^2}
+16\left(- 16\zeta_{\rm R}(3)+\zeta_{\rm R}(5)\right)\frac{1}{\Om^3}
+\dots \right)\,.
\ee
The asymptotic expansion for small argument is more elaborate and we represent it in the Appendix A, Eq. \Ref{A.2}, where one needs to put $r_0=1$ with coefficients given by Eq. \Ref{A.r1}.

It must be mentioned that the above derivation of $\F_1$ does not depend on which of the two aspects mentioned above, is considered. Moreover, it holds also in the high temperature limit which is, in the leading order, for all models given by the zeroth Matsubara frequency,
\be\label{4.4.2.Tinf}\F \raisebox{-4pt}{$\sim\atop T\to\infty$}
                    \frac{T}{4\pi^2}\,\phi(0),
\ee
which vanishes for the TE contribution to the Drude model whereas it is the same in both models for the TM polarization. In this way, the TE contribution is missing in the Drude model which, for example at large separation, where both polarization give the equal ideal conductor contributions to the plasma model, amounts for as much as the half of the free energy.

The {\bf second way} to derive $\F_1$ starts after the application of the Abel-Plana formula, i.e., from Eqs. \Ref{3.F3} and \Ref{3.F4} with the permittivity $\ep^{\rm Dr}(-\om)$, Eq. \Ref{4.4.1.ep}, inserted. The reflection coefficients are
\bea\label{4.4.2.epom}\rE^{\rm Dr}(-\om,\ga)&=&
        \frac{\eta-\sqrt{-\Om^2\frac{\om}{\om-i\ga}+\eta^2}}
                {\eta+\sqrt{-\Om^2\frac{\om}{\om-i\ga}+\eta^2}},
\nn\\
\rM^{\rm Dr}(-\om,\ga)&=&
        \frac{\left(1-\frac{\Om^2}{\om(\om-i\ga)}\right)\eta-\sqrt{-\Om^2\frac{\om}{\om-i\ga}+\eta^2}}
                {\left(1-\frac{\Om^2}{\om(\om-i\ga)}\right)\eta+\sqrt{-\Om^2\frac{\om}{\om-i\ga}+\eta^2}},
\eea
where we also indicated the dependence on $\ga$ explicitly. We remind, that for $\ga\to0$ these coefficients turn into that of the plasma model. Thus, a difference between both models may follow only if this limit and the integrations do not commute. Further we mention that both expressions, $ \F^{\rm Dr}$ and $ \F^{\rm pl}$, do exist. Both can be obtained and coincide if  carrying out formally the limit in the integrand. However, as known, this is a necessary, but not a sufficient condition. In fact, one needs a uniform convergence of the integrand, which is not there. Namely, this limit is not uniform for the TE-contribution,
\bea\label{4.4.2.mm}\lim_{\ga\to0}\rE^{\rm Dr}(\om,\ga)&=&\r(\Omega),   \nn\\
                    \lim_{\om\to0}\rE^{\rm Dr}(\om,\ga)&=&0.
\eea
In fact, this is the same behavior as observed in \Ref{4.4.2.r0} in the Matsubara representation.

Within the representation \Ref{3.F4} of the free energy, the non-commutativity of the limits \Ref{4.4.2.mm} shows up in the TE polarization in region \b, defined in \Ref{3.reg}. In region \a, from the integration volume, there is an additional factor $\om^2$ which ensures the commutativity of the limits. Thus we are faced with region \b,
\be\label{4.4.2.FE}\Delta_T\F_{\b\, \rm TE}=
    \frac{1}{4\pi^2}\int_0^\infty d\om\, \frac{1}{e^{\om/T}-1}
    i(\phi(i\om)-c.c.),
\ee
where
\be\label{4.4.2.fi}\phi(i\om)=\int_0^\infty
        d\eta\,\eta\,\ln\left(1-\left({\rE^{\rm Dr}(-\om,\ga)}\right)^2\,e^{-\eta}\right),
\ee
and we used \Ref{4.4.2.Om1}.

Now, in order to account for \Ref{4.4.2.mm}, we split the integration region in \Ref{4.4.2.FE} into two parts,
\be\label{4.4.2.AB}\Delta_T\F_{\b\, \rm TE}=A+B,
\ee
with
\be\label{4.4.2.A}A=    \frac{1}{4\pi^2} \int_0^{1} d\om\, \frac{1}{e^{\om/T}-1}\,
    i(\phi(i\om)-c.c.)
\ee
and
\be\label{4.4.2.B}B=\frac{1}{4\pi^2}\int_{1}^\infty
    d\om\, \frac{1}{e^{\om/T}-1}\,
    i(\phi(i\om)-c.c.).
\ee
In part A  we substitute $\om\to\om\ga$,
\be\label{4.4.2.A1}A=    \frac{\ga}{4\pi^2} \int_0^{1/\ga} d\om\, \frac{1}{e^{\om\ga/T}-1}\,
    i(\psi(i\om)-c.c.)
\ee
with
\bea\label{4.4.2.A3}\psi(i\om)&\equiv&\phi(i\om \ga)
\nn\\&=&    \int_0^\infty d\eta\,\eta\,\ln\left(1-\r\left(\Om\sqrt{\frac{\om}{\om-i}},\eta\right)^2\,e^{-\eta}\right)
\eea
which, in fact, does not depend on $\ga$. We used with \Ref{4.4.2.epom}
\be\label{4.4.2.r1}\rE^{\rm Dr}(\om,\om\ga)=\r\left(\Om \sqrt{\frac{\om}{\om-i}},\eta\right)
\ee
with $\r(\Omega,\eta)$ defined in \Ref{4.4.2.Om}. After this substitution we can put $\ga=0$ in the integrand and in the upper integration limit,
or tend $T\to0$ assuming a temperature dependent $\ga(T)$ with $\lim_{T\to0}\frac{\ga(T)}{T}=0$, in \Ref{4.4.2.A1} and obtain
\be\label{4.4.2.A2}A(\ga\to0)=    \frac{T}{4\pi^2} \int_0^{\infty} \frac{d\om}{\om}\,
    i(\psi(i\om)-c.c.).
\ee
In part B we can put $\ga=0$ directly in the integrand and in the lower integration limit and obtain the corresponding contribution to the plasma model.

In this way, we re-obtain Eq. \Ref{4.4.2.F1} with the right hand side given by A, Eq. \Ref{4.4.2.A2}. The last step is to carry out the integration over $\om$ in Eq. \Ref{4.4.2.A2}. This is possible using Eq. \Ref{B.4} accounting for $\phi(0)=0$, which holds for the function \Ref{4.4.2.A3}, and we get
\be\label{4.4.2.A0}A(\ga\to0)=\F_1\,T
\ee
with $\F_1$ given by Eq. \Ref{4.4.2.F3}.

In fact, the division \Ref{4.4.2.AB} of $\Delta_T\F_{\b\, \rm TE}$ was observed in \cite{bord11-71-1788} in an attempt to pass from the Matsubara summation to an integration over real frequencies, just the same way as one does applying the Abel-Plana formula. In \cite{bord11-71-1788} it was shown what happens to the integration path under deformation, $\xi=\om e^{-i \beta}$, with $\beta$ changing from $\beta=0$ to $\beta=\pi/2$. It was found (see  \cite{bord11-71-1788}, p.9) that this path with increasing $\beta$ develops a self-intersection and that, for $\beta=\pi/2$ in the limit of $\ga\to0$, it decays into two separate pathes which just correspond to the two contributions, $A$ and $B$, in \Ref{4.4.2.AB}. In \cite{bord11-71-1788} this contribution was found analytically.

\subsection{Insulator described by oscillator model}
For an insulator, we use the permittivity given by Eq. \Ref{2.5.ep} and for the low temperature expansion we need to insert
\be\label{4.5.ep}\ep^{\rm insul.}(-\om)=1+\sum_j\frac{g_j}{\om_j^2-\om^2+i\ga_j\om}
\ee
into $\phi(i\om)$, Eq. \Ref{3.f22}. Like in Sect. 4.2, with the permittivity \Ref{4.5.ep}, the free energy of an ideal conductor does not follow in any limit. Therefor we use in this section also the four cases, introduced with Table \ref{4.2.cases}.

We start with the special case of no dissipation, $\ga_j=0$ in \Ref{4.5.ep}. This is a real permittivity. We expand it for small $\om$,
\be\label{4.5.ep1}\ep^{\rm insul.}(-\om)=\ep_0+\frac{\om^2}{\om_0^2}+\dots
\ee
with
\be\label{4.5.ep0}\ep_0=1+\sum_j\frac{g_j}{\om_j^2},\qquad
        \frac{1}{\om_0^2}=\sum_j\frac{g_j}{\om_j^4}.
\ee
The absence of an imaginary part in \Ref{4.5.ep1} allows to use the results obtained in Sect. 4.2 for fixed permittivity and simply to insert \Ref{4.5.ep1} into \Ref{4.2.3.results}, \Ref{4.2.3.ie} and \Ref{4.2.3.ee} with subsequent  re-expansion in powers of $\om$. Since the expansion in \Ref{4.5.ep1} is in even powers of $\om$, the first correction due to \Ref{4.5.ep1} occurs in the order $\om^4$, and the next in order $\om^5$. Denoting the coefficients of the expansion according to Eq. \Ref{3.Fi2}, one has to substitute, besides $\ep\to\ep_0$, in the coefficients
\bea\label{4.5.subst}\Phi_{(k),4}&\to&\Phi_{(k),4}+\frac{1}{\om_0^2}\Phi_{(k),2},
 \nn\\              \Phi_{(k),5}&\to&\Phi_{(k),5}+\frac{1}{\om_0^2}\Phi_{(k),2},
\eea
where $(k)$ denotes the case according to Table \ref{4.2.cases}.

Now we allow for dissipation, $\ga_j\ne0$.
In this case the expansion of the permittivity is
\be\label{4.5.ep2}\ep^{\rm insul.}(-\om)=\ep_0-i\ga_0\om+\dots
\ee
with
\be\label{4.5.ga}\ga_0= \sum_j\frac{g_j\ga_j}{\om_j^4}.
\ee
Cross terms, involving corrections from both, \Ref{4.5.ep1} and \Ref{4.5.ep2}, do not occur in the orders which we are interested in.

We divide the integration in $\phi(i\om)$ into the regions \Ref{3.reg}. In region \a\, we use the same formula, Eq. \Ref{4.2.1.fi} as in Sect. 4.2. We can go the same way as there and obtain an expansion which starts from $\om^3$ which will be non leading.
The leading contribution comes from region \b. Here we have to use Eq. \Ref{3.f22},
\be\label{4.5.fi}\phi_\b (i\om)=
\int_0^\infty d\eta\,\eta\,\ln\left(1-r_1r_2e^{-\eta}\right),
\ee
with the reflection coefficients \Ref{2.2.reta} to be inserted according to the cases in Table \ref{4.2.cases}. As it would be too elaborate and of less use to get higher orders of the expansion for small $\om$, we restrict ourselves here to the leading order, which is $\om$, and to the logarithmic contribution in order $\om^3$, i.e., the contribution being proportional to $\ln\om\,\om^3$, which is easy to obtain. We do not calculate the contributions proportional to $\om^3$ without logarithm. The simple reason for this restriction is in the structure of the integrand in  Eq. \Ref{4.5.fi}, where an expansion in powers of $\om$ delivers powers of $\eta$ in the denominator which make the integration over $\eta$ divergent. Also, a substitution $\eta\to\eta\om$ is not helpful since it produces a problem on the upper interaction boundary. The leading order can be obtained from expanding the logarithm in \Ref{4.5.fi} in powers of $\om$,
\be\label{4.5.ln}\ln\left(1-r_1r_2e^{-\eta}\right)=f_1(\eta)\om+f_3(\eta)\om^3+\dots\,.
\ee
We consider here odd powers of $\om$ only since the even powers will drop out later from the imaginary part. The function $f_1(\eta)$ is integrable, whereas the function $f_3(\eta)$ behaves as
\be\label{4.5.f3}f_3(\eta)=\frac{f_{30}}{\eta^2}+O(1)
\ee
for $\eta\to0$. This behavior results in a logarithmic singularity when inserted into \Ref{4.5.fi}. As a consequence, the expansion of the function $\Phi(\om)$ has a logarithmic contribution,
\be\label{4.5.Fi}\Phi_{(k)}(\om)=
    \Phi_{(k),1}\om+\left(\Phi_{(k),3}+\tilde{\Phi}_{(k),3}\right)\om^3+O(\om^5)
\ee
where we used the notation of Eq. \Ref{3.Fi2}. The coefficients $ \Phi_{(k),1}$ can be obtained by simple integration from $f_1(\eta)$ for a specific case according to Table \ref{4.2.cases},
\be\label{4.5.F1} \Phi_{(k),1}=\int_0^\infty d\eta\,\eta \,i(f_1(\eta)-\cc).
\ee
For the coefficient $\tilde{\Phi}_{(k),3}$ we split the integration,
\be\label{4.5.div} \int_0^\infty d\eta\,\eta \,i(f_3(\eta)-\cc)
=\int_0^\om d\eta\,\eta \,i(f_3(\eta)-\cc)
+\int_\om^\infty d\eta\,\eta \,i(f_3(\eta)-\cc).
\ee
The leading contribution comes from the lower integration limit in the second term of the right side and we get
\be\label{4.5.F3}\tilde{\Phi}_{(k),3}=-i(f_{30}-\cc).
\ee
As said already, we do not calculate ${\Phi}_{(k),3}$. The results are shown in Table \ref{4.5.tab}. These contributions are all due to dissipation. In order to get the corresponding results for the \elm case, the polarizations must be added, cases 1 and 3 for an ideal conductor in front of an insulator and cases 2 and 4 for two insulators.

\begin{table}[h]
  \centering
\begin{tabular}{c|r|r}
case    &   $\Phi_{(k),1}$  &   $\tilde{\Phi}_{(k),3}$  \\\hline &&\\[-6pt]
1   &   0   &   $\frac{1}{2}\,\ga_0$\\[4pt]
2   &   0   &   0\\[4pt]
3   &   $-\frac{4}{\ep_0^2-1} {\rm Li_2}\left(\frac{\ep_0-1}{\ep_0+1}\right)\,\ga_0$
        &  $ -\frac{(\ep_0^2+2\ep_0-1)}{(\ep_0+1)^2}\,\ga_0$\\[4pt]
4   &   $-\frac{8}{\ep_0^2-1} {\rm Li_2}\left(\frac{\ep_0-1}{\ep_0+1}\right)\,\ga_0$
        &  $ -\frac{(\ep_0-1)(\ep_0+3)}{(\ep_0+1)^2}\,\ga_0$
\end{tabular}
  \caption{The coefficients $\Phi_{(k),1}$ and $\tilde{\Phi}_{(k),3}$ calculated in Sect. 4.5 for an insulator with dissipation.}\label{4.5.tab}
\end{table}

\subsection{The case of  dc conductivity}
In this subsection we investigate the contribution of a dc conductivity to the free energy. For imaginary frequency, the permittivity is given by
\be\label{4.6.ep}\ep^{\rm dc}(i\xi)=\ep_0+\frac{4\pi\sigma}{\xi},
\ee
assuming $\ep_0\ne 1$. The behavior of this model is quite similar to the Drude model considered in Sect. 4.4. Here, it is the conductivity $\sigma$, which describes the dissipation of energy. The counterpart, in which the free energy corresponding to \Ref{4.6.ep} does not turn, is the insulator considered in the preceding subsection in place of the plasma model in Sect. 4.4. In the following subsections we consider separately the cases of non vanishing and of vanishing conductivity $\sigma$.

\subsubsection{Fixed conductivity $\sigma$}
The case of a  fixed conductivity $\sigma$, not depending on temperature, is in fact nonphysical and of rather academic interest. We include it here since the calculations go in parallel to a fixed dissipation parameter $\ga$ in the Drude model, only with the role of the two polarizations interchanged.  In both cases the permittivity behaves   the same way for small frequency, it has $\om$ in the denominator to the first power. The difference between both is in the behavior for large frequency, where $\ep^{\rm dc}(\om)\to\ep_0$ with $\ep_0\ne 1$, whereas $\ep^{\rm Dr}(\om)\to 1$ in this limit.

For the calculation we use the division into cases, given by Table \ref{4.2.cases}, and into regions \Ref{3.reg}. We are interested in the leading order only which is $\om$, including $\om\ln\om$. The region \a\, gives   contributions starting from $\om^2$, which we do not consider. Following \Ref{3.f22}, in region \b\, we have to consider
\be\label{4.6.1.fi}\phi_{(k)}(i\om)=\int_0^\infty d\eta\,\eta\,\ln\left(1-r_1r_2e^{-\eta}\right)
\ee
with $r_1$ and $r_2$ to be substituted according to Table \ref{4.2.cases} by one of the reflection coefficients from \Ref{2.2.reta} with the permittivity
\be\label{4.6.1.ep}\ep^{\rm dc}(-\om)=\ep_0+\frac{4\pi\sigma}{i\om}.
\ee
and $\kappa$ from Eq. \Ref{2.2.kappa}.

The further considerations are in parallel to the Drude model, Sect. 4.4., and we do not repeat all discussions. For the case 1, for    the function $\phi_{(1)}(i\om)$, we use Eq. \Ref{A.1}, where we have to substitute
\be\label{4.6.1.A}r_0\to -1,\qquad \Omega^2\to 4\pi i \sigma \om -(\ep_0-1)\om^2.
\ee
From the leading order coefficient from Eq. \Ref{A.r-1} and taking the imaginary part according to Eq. \Ref{3.Fi}, we get
\be\label{4.6.1.F1}\Phi_{(1)}(\om)=2\pi\sigma\left(-1+\gamma_{\rm E}+\frac12\ln(16\pi\sigma\om)\right)\om+O(\om^{3/2}),
\ee
which has a logarithmic contribution.

Case 2 is easier. After substituting $\eta\to\sqrt{\om}\eta$ in \Ref{4.6.1.fi}, one can expand the integrand,
\be\label{4.6.1.fi2}\phi_{(2)}(i\om)=\int_0^\infty d\eta\,\eta\,
    \ln\frac{4\eta\sqrt{\eta^2+4\pi p \sigma}}
    {(\eta+\sqrt{\eta^2+4\pi i \sigma})^2}      +O(\om^{3/2})
\ee
with subsequent integration. This results in
\be\label{4.6.1.F2}\Phi_{(2)}(\om)= 2\pi\sigma(\ln 2-1)\om+O(\om^{3/2}).
\ee
In the leading order, considered here, the results are in parallel to the Drude case, considered in Sect. 4.4.1. This can be seen comparing the substitutions \Ref{4.4.1.A} and \Ref{4.6.1.A}. To leading order these are connected by the substitution $\frac{\Om^2}{\ga}\to 4\pi\sigma$. This holds also for the next-to-leading order, which is proportional to $\om^{3/2}$ and which results from the coefficient $a_3$ from the Appendix A. In this way, the formulas \Ref{4.4.1.Fi2n} and \Ref{4.4.1.Fi1n} from the Drude model translate by the above transformation into the corresponding formulas for fixed dc conductivity. It is seen this way that there is no dependence on $\ep_0$ in the considered orders of the expansion.

In the cases 3 and 4, i.e., for the TM polarization, one expands the integrand directly for small $\om$,
\be\label{4.6.1.fi3}\phi_{(3)}(i\om)=
    \int_0^\infty d\eta\,\frac{\eta}{2\pi i \sigma(e^\eta -1)}    +O(\om^{2}),
\ee
which results in
\be\label{4.6.1.F3}\Phi_{(3)}(\om)= -\frac{\pi}{6\sigma}\om+O(\om^{ 2}).
\ee
In case 4 the result is just twice,
\be\label{4.6.1.F4}\Phi_{(4)}(\om)= -\frac{\pi}{3\sigma}\om+O(\om^{ 2}).
\ee
These results are similar to that of the Drude model, Sect. 4.4.1, and can be obtained from \Ref{4.4.1.Fi3} and \Ref{4.4.1.Fi4}  by the above substitution.

\subsubsection{Vanishing conductivity $\sigma$}
In this subsection we consider the free energy for vanishing dissipation parameter, $\sigma\to0$. As before in the Drude model there are two aspects. The first is the entropy $S$, Eq. \Ref{1.S}, having a residual value
\be\label{4.6.2.S0}\lim_{T\to0} S=-\F_1^\sigma
\ee
similar to \Ref{4.4.2.S0} for vanishing temperature in case $\sigma$ goes to zero not slower than the first power of the temperature such that
\be\label{4.6.2.lim}\lim_{T\to0}\frac{\sigma(T)}{T}=0
\ee
holds. The second aspect is that the free energy with dc conductivity, $\F^{\rm dc}$, for $\sigma\to0$ at fixed temperature, does not reproduce that of the insulator $\F^{\rm insul.}$ considered in Sect. 4.5.,
\be\label{4.6.2.F1}\F^{\rm dc}\raisebox{-4pt}{$\sim\atop T\to0$}\F^{\rm insul.}
        +\F_1^\sigma T,
\ee
where $\F_1^\sigma$ is the same   as in \Ref{4.6.2.S0}. The comments made on this in the Drude model apply here too.

There are two ways to show the above statement and to calculate $\F_1^\sigma$. The first way uses the Matsubara representation. The reflection coefficients are given by Eq. \Ref{2.2.reta} with \Ref{2.2.kappa} and the permittivity \Ref{4.6.ep},
\bea\label{4.6.2.r}\rE^{\rm dc}(\xi_l,\sigma)
    &=&\frac{\eta-\sqrt{\ep_0\xi_l^2+4\pi\sigma\xi_l+\eta^2}}
            {\eta+\sqrt{\ep_0\xi_l^2+4\pi\sigma\xi_l+\eta^2}}
 ,\nn\\ \rM^{\rm dc}(\xi_l,\sigma)
    &=&\frac{\left(\ep_0+\frac{4\pi\sigma}{\xi_l}\right)\eta-\sqrt{\ep_0\xi_l^2+4\pi\sigma\xi_l+\eta^2}}
            {\left(\ep_0+\frac{4\pi\sigma}{\xi_l}\right)\eta+\sqrt{\ep_0\xi_l^2+4\pi\sigma\xi_l+\eta^2}},
\eea
and the corresponding coefficients for the insulator are given by the above for $\sigma=0$,
\bea\label{4.6.2.r1}\rE^{\rm insul.}(\xi_l )
    &=&\frac{\eta-\sqrt{\ep_0\xi_l^2 +\eta^2}}
            {\eta+\sqrt{\ep_0\xi_l^2 +\eta^2}}
 ,\nn\\ \rM^{\rm insul.}(\xi_l )
    &=&\frac{\ep_0\eta-\sqrt{\ep_0\xi_l^2 +\eta^2}}
            {\ep_0\eta+\sqrt{\ep_0\xi_l^2 +\eta^2}}
\eea
for all $l$. Now,  there is a difference between both for zero  Matsubara frequency, $l=0$. For dc conductivity we have
\be\label{4.6.2.r0}\rE^{\rm dc}(0,\sigma)
   =0,\qquad \rM^{\rm dc}(0,\sigma)
   =1,
\ee
whereas, for the insulator,
\be\label{4.6.2.r10}\rE^{\rm insul.}(0 )
   =0,\qquad \rM^{\rm insul.}(0 )
   =\frac{\ep_0-1}{\ep_0+1},
\ee
hold. As compared to the Drude model, in this case the TM polarization delivers the difference. In all other contributions the limit $\sigma\to0$ is smooth  and we get
\be\label{4.6.2.dF}
\lim_{\sigma\to0}\left( \F^{\rm dc}-\F^{\rm insul.}\right)=\F_1^\sigma T
\ee
with
\be\label{4.6.2.Fs}\F_1^\sigma=
\frac{1}{4\pi^2}\int_0^\infty d\eta\,\eta
\left(  \ln\left(1-e^{-\eta}\right)- \ln\left(1-r_0^2e^{-\eta}\right)\right)
\ee
with
\be\label{4.6.2.r00}r_0=\frac{\ep_0-1}{\ep_0+1}.
\ee
The integration can be carried out delivering
\be\label{4.6.2.Fs1}\F_1^\sigma=\frac{1}{4\pi^2}\left(\zeta_{\rm R}(3)-{\rm Li}_3(r_0^2)\right)
\ee
(see \cite{geye05-72-085009}), which comes in place of \Ref{4.4.2.F3} in the Drude model.

The second way to derive $\F_1^\sigma$ starts from representation \Ref{3.F3} with \Ref{3.Fi} and the permittivity \Ref{4.6.ep}. The reflection coefficients are
\bea\label{4.6.2.rom}\rE^{\rm dc}(-\om,\sigma)&=&
    \frac{\eta-\sqrt{\ep_0\om^2-4\pi i\sigma\om+\eta^2}}
            {\eta+\sqrt{\ep_0\om^2-4\pi i\sigma\om+\eta^2}}.
\nn\\   \rM^{\rm dc}(-\om,\sigma)&=&
    \frac{\left(\ep_0+\frac{4\pi i\sigma}{\om}\right)\eta-\sqrt{\ep_0\om^2-4\pi i\sigma\om+\eta^2}}
            {\left(\ep_0+\frac{4\pi i\sigma}{\om}\right)\eta+\sqrt{\ep_0\om^2-4\pi i\sigma\om+\eta^2}}.
\eea
The coefficient of the TM polarization does not have a   limit for $\sigma\to0$ which would be uniform in $\om$. Again, the interesting contribution comes from region \b,
\be\label{4.6.2.DF}\Delta_T\F^{\rm dc}_{\rm TM}=\frac{1}{4\pi^2}\int_0^\infty d\eta\,\eta
\ln\left(1-\left(\rE^{\rm dc}(-\om,\sigma)\right)^2e^{-\eta}\right),
\ee
which we split also,
\be\label{4.6.2.AB}\Delta_T\F^{\rm dc}_{\rm TM}=A+B,
\ee
with
\be\label{4.6.2.A}A=    \frac{1}{4\pi^2} \int_0^{1} d\om\, \frac{1}{e^{\om/T}-1}\,
    i(\phi(i\om)-c.c.)
\ee
and
\be\label{4.6.2.B}B=\frac{1}{4\pi^2}\int_{1}^\infty
    d\om\, \frac{1}{e^{\om/T}-1}\,
    i(\phi(i\om)-c.c.),
\ee
The part B turns, for $\sigma\to0$ into the corresponding one of the insulator. In A we substitute $\om\to\om \sigma$,
\be\label{4.6.2.A1}A=    \frac{\sigma}{4\pi^2} \int_0^{1/\sigma} d\om\, \frac{1}{e^{\om\sigma/T}-1}\,
    i(\psi(i\om)-c.c.)
\ee
with
\bea\label{4.6.2.psi}\psi(i\om)&\equiv&\phi(i\om \sigma),
\nn\\&=&    \int_0^\infty d\eta\,\eta\,\ln\left(1-
r_1\left(\ep_0+\frac{4\pi i}{\om},\eta\right)^2\,e^{-\eta}\right).
\eea
The function $r_1(\Omega,\eta)$ is given by Eq. \Ref{4.4.2.Om} as before and $\psi(i\om)$ is, in fact, independent on $\sigma$.

Now we can tend $\sigma\to0$ in \Ref{4.6.2.A1} and get
\be\label{4.6.2.A2}A(\sigma\to0)= \frac{T}{4\pi^2} \int_0^{\infty} \frac{d\om}{\om}\,
    i(\psi(i\om)-c.c.)
\ee
The last step is to apply Eq. \Ref{B.4} with $\phi(i\om)\to\psi(i\om)$ which results in
\be\label{4.6.2.A3}A(\sigma\to0)=\F_1^\sigma T
\ee
and finishes the second way to  derive  $\F_1^\sigma$.

\subsection{Hydrodynamic model for graphene}
The hydrodynamic model is described by the reflection coefficients \Ref{2.7.r}. Like in the plasma model in Sect. 4.3, we allow for different plasma frequencies, $\Omi$ $(i=1,2)$, in the two interfaces. The corresponding reflection coefficients are
\be\label{4.7.r}{\rE}_i=\frac{-1}{1-\frac{i q}{\Omi}},\qquad
                {\rM}_i=\frac{1}{1+\frac{\om^2}{i q \Omi}}.
\ee
For the calculation of the low temperature expansion we use again the division into the two regions \Ref{3.reg}.
\subsubsection{Region \a}
Here we have a real $q$ and the function \Ref{3.fq} is
\be\label{4.7.1.fiE}\phi_\a (i\om)=\int_0^\om dq\,q\,\ln\left(1-r_1r_2e^{-iq}\right),
\ee
where for $r_i$ one needs to insert ${\rE}_i$ or ${\rM}_i$ according to the polarization. In this model, the calculation of the expansion of for small $\om$ is not such easy as for the plasma model in Sect. 4.3.1, for instance, the reflection coefficients are not pure phase factors.

For the TE polarization, however, the expansion can be obtained easily since an expansion of the integrand in powers of $\om$ does not cause problems with the convergence of the subsequent integration over $q$. In this way one can obtain the expansion machinized and insert into \Ref{3.Fi}, and carrying out the integration the same way one obtains
\bea\label{4.7.1.FE}\Phi_{\a {\rm TE}}(\om)&=&-\frac{\pi}{2}\,\om^2+
\left(\frac13+\frac{\Oma^2 \Omb+2 \Oma^2+\Oma \Omb^2+2 \Oma
   \Omb+2 \Omb^2}{3 \Oma \Omb (\Oma
   \Omb+\Oma+\Omb)}\right)\,\om^3\nn\\&&+O(\om^5),
\eea
where we restricted ourselves to display the first two contributions only.

For the TM polarization such a simple expansion does not work since an expansion in powers of $\om$ would produce inverse powers of $q$ which make the integration divergent. We proceed as follows. We split the functions $\phi_{\a {\rm TM}}(i \om)$ into two parts,
\be\label{4.7.1.AB}\phi_{\a {\rm TM}}(i \om)=A+B,
\ee
with
\be\label{4.7.1.A}A=\int_0^\om dq\,q\,\ln\left(1-{\rM}_1 {\rM}_2\right)
\ee
and
\be\label{4.7.1.B}B= \int_0^\om dq\,q\,
        \ln\left(1-\frac{e^{-iq}-1}{({\rM}_1 {\rM}_2)^{-1}-1}\right).
\ee
The integration in $A$ can be carried out explicitly with subsequent expansion in powers of $\om$.
It should be mentioned, that the resulting expression does not depend on the separation $a$ which enters, when restoring the dimensional parameters as discussed in sect.2, only through the exponential, $\exp(-2ia\eta)$.

In $B$, the integrand can be expanded up to $\om^6$ with convergent subsequent integration. Adding the results from $A$ and $B$ and using \Ref{3.Fi} one obtains the expansion
\bea\label{4.7.1.FM}\Phi_{\a {\rm TM}}(\om)&=&-\frac{\pi}{2}\,\om^2
+\left(\frac{1}{3}+\frac{1}{\Oma}+\frac{1}{\Omb}
+\frac{\left(\Oma^2+\Omb^2\right)
\arctan\left(\sqrt{\frac{\Oma  \Omb}{\Oma+\Omb}}\right)       }
{(\Oma \Omb)^{3/2}
   \sqrt{\Oma+\Omb}}
   \right)\,\om^3
\nn\\&&     -\frac{\pi  \left(\Oma^4+2 \Oma^3 \Omb+\Oma^2
   \Omb^2+2 \Oma \Omb^3+\Omb^4\right)}{2 \Oma^2
   \Omb^2 (\Oma+\Omb)^2}\,\om^4+O(\om^5),
\eea
which completes the contributions from region \a.

\subsubsection{Region \b}
In this reagin we have a real $\eta$ and, inserting $q=i\eta$ in \Ref{4.7.r}, real reflection coefficients, which must be inserted into \Ref{4.3.2.fi}. Thus, just as in the plasma model, we have a seemingly real function $\phi_\b(i \om)$, Eq. \Ref{3.f22}. It can have an imaginary part only from the logarithm when its argument changes sign. Indeed, the hydrodynamic model, in parallel to the plasma model, is known to have surface plasmons in the TM polarization. In order to account for them we rewrite Eq. \Ref{3.f22},
\be\label{4.7.2.fi}\phi_\b(i \om)=\int_0^\infty d\eta\,\eta\left(
-\ln({\rM}_1)^{-1}-\ln({\rM}_2)^{-1}+\ln\left(({\rM}_1)^{-1}({\rM}_2)^{-1}-e^{-\eta}\right)
\right),
\ee
removing the poles from the arguments of the logarithms and separating the contributions from the plasmons  on the interfaces taken alone, $\eta_{{\rm single},i}$ $(i=1,2)$, which are solution of
\be\label{4.7.2.eqsi}({\rM}_i)^{-1}=0
\ee
and the two plasmons $\eta_{{\rm symm.}}$ and $\eta_{\rm antisy.}$ which are solutions of the equation
\be\label{4.7.2.eq2}({\rM}_1)^{-1}({\rM}_2)^{-1}=e^{-\eta}.
\ee
We took the notations in accordance with the symmetry the corresponding wave functions have in case of equal plasma frequencies.

With \Ref{4.7.r}, the solutions of equation \Ref{4.7.2.eqsi} are obviously
\be\label{4.7.2.etasi}\eta_{{\rm single},i}=\frac{\om^2}{\Omi}.
\ee
The equations \Ref{4.7.2.eq2} are transcendental ones like in the case of the plasma model, Sect. 4.3.2. There is only a small difference to the case of the plasma model as here the solution $\eta_{\rm antisy.}$ exists also for arbitrarily small $\om$.

The solution $\eta_{{\rm symm.}}$ can be obtained from the equation \Ref{4.7.2.eq2}. We solve this equation for $\eta$ occurring in the left side, which amounts in solving a quadratic equation. With the appropriate one of the two solutions we rewrite the equation in the form
\be\label{4.7.2.eq3}\eta=\frac
    {1-\sqrt{1- \mu\left(1-e^{-\eta}\right)}}{\mu\left(1-e^{-\eta}\right)}
    \frac{\om^2}{\Oma+\Omb},
\ee
where we introduced the notation
\be\label{4.7.2.mu}\mu\equiv \frac{2\Oma \Omb}{(\Oma+\Omb)^2}.
\ee
This equation can be solved by iteration, starting from inserting $\eta=0$ in the right side. After re-expanding in powers of $\om$, the solution is
\be\label{4.7.2.etasy}\eta_{{\rm symm.}}=
 \frac{\om^2}{\Oma+\Omb}+\frac{\mu}{2}   \left( \frac{\om^2}{\Oma+\Omb}\right)^2
 +\frac{\mu(3\mu-1)}{4}\left( \frac{\om^2}{\Oma+\Omb}\right)^3+O(\om^7).
\ee
In order to obtain the solution $\eta_{\rm antisy.}$, one needs to take the other solution of the quadratic equation mentioned above and one needs to rewrite the equation in the form
\be\label{4.7.2.eq4}\eta=
\sqrt{\frac{\eta}{1-e^{-\eta}}\,\frac{1+\sqrt{1-2\mu\left(1-e^{-\eta}\right)}}{2}}
                \sqrt{\frac{\Oma+\Omb}{\Oma \Omb}}\,\om.
\ee
Iteration, starting from inserting $\eta=0$ in the right hand side, gives
\bea\label{4.7.2.etaas}\eta_{\rm antisy.}&=&
\sqrt{\frac{\Oma+\Omb}{\Oma \Omb}}\,\om
+\frac{1-\mu}{4}\left(\sqrt{\frac{\Oma+\Omb}{\Oma \Omb}}\,\om\right)^2
\nn\\&&+\frac{7-7\mu-9\mu^2}{96}\left(\sqrt{\frac{\Oma+\Omb}{\Oma \Omb}}\,\om\right)^3
\nn\\&&+\frac{1-\mu-3\mu^2}{48}\left(\sqrt{\frac{\Oma+\Omb}{\Oma \Omb}}\,\om\right)^4
+O(\om^5).
\eea
Using these solutions, we return to Eq. \Ref{4.7.2.fi}. The analytic continuation can be carried out starting from large $\eta$, where the logarithms are real. The arguments of the logarithms change sign whenever, for decreasing $\eta$, these pass a zero, and the logarithms acquire an addendum $i\pi$. The argument of the last logarithm in \Ref{4.7.2.fi} has two zeros, $\eta_{\rm antisy.}$ and $\eta_{\rm symm.}$, whereby the inequality
\be\label{4.7.2.ineq}\eta_{\rm symm.}<\eta_{\rm antisy.}
\ee
holds, which can be verified for small $\om$ with the leading terms in \Ref{4.7.2.etasy} and \Ref{4.7.2.etaas}. Since there is no pole in between, the logarithm acquires an addendum of $-i\pi$ in passing $\eta_{\rm symm.}$, i.e., the second zero. In this way we get from \Ref{4.7.2.fi}, and accounting for \Ref{3.Fi},
\be\label{4.7.2.}\Phi_\b(\om)=-2\pi\left(
-\int_0^{\eta_{{\rm single},1}} d\eta\,\eta
-\int_0^{\eta_{{\rm single},2}} d\eta\,\eta
+\int_{\eta_{\rm symm.}}^{\eta_{\rm antisy.} } d\eta\,\eta        \right).
\ee
Carrying out the integrations and re-expanding for small $\om$, we get
\bea\label{4.7.2.F2M}\Phi_\b(\om)&=&
-\frac{\pi}{2}\left(\frac{1}{\Oma}+\frac{1}{\Omb}\right)\,\om^2
+-\frac{\pi  \left(\Oma^2+\Omb^2\right)}{4 \sqrt{2} \sqrt{\Oma^3 \Omb^3
   (\Oma+\Omb)}}\,\om^3
\\\nn&&+\frac{\pi  \left(91 \Oma^4+184 \Oma^3 \Omb+294 \Oma^2 \Omb^2+184 \Oma
   \Omb^3+91 \Omb^4\right)}{96 \Oma^2 \Omb^2 (\Oma+\Omb)^2}\,\om^4+O(\om^5).
\eea
This is the complete contribution from region \b\, since the TE polarization does not have surface plasmons. The two first leading order contributions come from the antisymmetric plasmon.

\subsubsection{The low frequency expansion}
Here we collect the results from the preceding two subsections. For the TE polarization we have only the contribution from region \a, which is given by Eq. \Ref{4.7.1.FE}
\be\label{4.7.3.TE}\Phi_{\rm TE}(\om)=\Phi_{\a {\rm TE}}(\om).
\ee
Special cases are equal plasma frequencies,
\bea\label{4.7.3.Eeq}{\Phi_{\rm TE}(\om)}_{{ |}_{\Oma=\Omb=\Om}} &=&
-\frac{\pi}{2}\,\om^2
+\frac{2 (\Om+3)}{3 \Om (\Om+2)} \,\om^3
\\\nn&&-\frac{\Om^4+12 \Om^3+54 \Om^2+96 \Om+60}{30 \Om^3
   (\Om+2)^3} \,\om^5
+O(\om^7),
\eea
and one interface ideal conducting,
\bea\label{4.7.3.Eid}{\Phi_{\rm TE}(\om)}_{{ |}_{\Oma=\infty,\Omb=\Om}} &=&
-\frac{\pi}{2}\,\om^2
+ \frac{\Om+2}{3 \Om( \Om+1)}\,\om^3
\\\nn&&-\frac{\Om^4+8 \Om^3+24 \Om^2+24 \Om+8}{60 \Om^3
   (\Om+1)^3}\,\om^5
+O(\om^7).
\eea
For the TM polarization we have to add \Ref{4.7.1.FM} and \Ref{4.7.2.F2M},
\bea\label{4.7.3.TM}{\Phi_{\rm TM}}(\om)&=&
    -\frac{\pi}{2}\left(1+\frac{1}{\Oma}+\frac{1}{\Omb}\right)\,\om^2
\nn\\\nn&&    +\left(
   \frac13+\frac{1}{\Oma}+\frac{1}{\Omb}
    -\frac{\pi  \left(\Oma^2+\Omb^2\right)}{4 \sqrt{2} \sqrt{\Oma^3
   \Omb^3 (\Oma+\Omb)}}
\right.\nn\\&&\left.
+\frac{\left(\Oma^2+\Omb^2\right) \arctan\left(\sqrt{\frac{\Oma  \Omb}
   {\Oma+\Omb}}\right)}{(\Oma \Omb)^{3/2}
   \sqrt{\Oma+\Omb}}
                        \right)\,\om^3
\\\nn&&+\frac{\pi  \left(43 \Oma^4+88 \Oma^3 \Omb+246 \Oma^2
   \Omb^2+88 \Oma \Omb^3+43 \Omb^4\right)}{96 \Oma^2
   \Omb^2 (\Oma+\Omb)^2}    \,\om^4+O(\om^5).
\eea
Special cases are, again, equal plasma frequencies,
\bea\label{4.7.3.Meq}{\Phi_{\rm TM}(\om)}_{{ |}_{\Oma=\Omb=\Om}} &=&
-\frac{\pi}{2}\left(1+\frac{2}{\Om}\right)\,\om^2
+\left(\frac{1}{3} +\frac{2}{\Om}-\frac{\pi }{4 \Om^{3/2}}
        +\frac{\sqrt{2} \arctan  \left(\frac{\sqrt{\Om}}{\sqrt{2}}\right)}{\Om^{3/2}}
            \right) \,\om^3
\nn\\&&
                +\frac{127\pi}{96\Om^2}\,\om^4
+\left(
    \frac{-16 \Om^3+47 \Om^2+402 \Om+480}{96 \Om^3
   (\Om+2)^2}
\right.\\\nn&&\left.   %
-\frac{5 \pi }{512 \Om^{5/2}}
+\frac{5 \arctan\left(\frac{\sqrt{\Om}}{\sqrt{2}}\right)}{32 \sqrt{2}
   \Om^{5/2}}       \right) \,\om^5
   +\frac{701 \pi }{11520 \Om^3}\,\om^6
+O(\om^7),
\eea
and one interface ideal conducting,
\bea\label{4.7.3.Mid}{\Phi_{\rm TM}(\om)}_{{ |}_{\Oma=\infty,\Omb=\Om}} &=&
-\frac{\pi}{2}\,\om^2
+ \left(\frac   {1}{3}+\frac{1}{\Om} -\frac{\pi }{4 \sqrt{2} \Om^{3/2}}+\frac{\arctan
   \left(\sqrt{\Om}\right)}{\Om^{3/2}}\right)\,\om^3
\nn \\\nn&&  +\frac{43 \pi }{96 \Om^2}\,\om^4
+\left(
-\frac{5 \pi }{256 \sqrt{2} \Om^{5/2}}
\frac{-8 \Om^3+39 \Om^2+113 \Om+64}{96 \Om^3
   (\Om+1)^2}
\right.\\ &&\left. +\frac{5 \arctan \left(\sqrt{\Om}\right)}{512 \Om^{5/2}}
                     \right)\,\om^5
-\frac{19 \pi }{5760 \Om^3}\,\om^6+O(\om^7).
\eea
Finally we add the polarizations and get the expansion for the \elm case,
\bea\label{4.7.3.ED}{\Phi_{\rm ED}}(\om) &=&
-\left( 1+\frac{1}{2 \Oma}+\frac{1}{2 \Omb}  \right)\pi\,\om^2
\\\nn&&+\left(  \frac{2 \Oma^2 \Omb^2+6 \Oma^2 \Omb+5 \Oma^2+6
   \Oma \Omb^2+8 \Oma \Omb+5 \Omb^2}{3 \Oma
   \Omb (\Oma \Omb+\Oma+\Omb)}
\right.\nn\\\nn&&\left.%
-\frac{\pi  \left(\Oma^2+\Omb^2\right)}{4 \sqrt{2} \sqrt{\Oma^3
   \Omb^3 (\Oma+\Omb)}}
+\frac{\left(\Oma^2+\Omb^2\right) \arctan  \left(\sqrt{\frac{\Oma
   \Omb}{\Oma+\Omb}}\right)}{(\Oma \Omb)^{3/2}
   \sqrt{\Oma+\Omb}}\right)\,\om^3
\\\nn&&+\left(\frac{\pi  \left(43 \Oma^4+88 \Oma^3 \Omb+246 \Oma^2
   \Omb^2+88 \Oma \Omb^3+43 \Omb^4\right)}{96 \Oma^2
   \Omb^2 (\Oma+\Omb)^2}  \right)\,\om^4+O(\om^5).
\eea
Special cases are considered in the next section.
The coefficients of these expansions, identified according to Eq. \Ref{3.Fi2}, must be inserted into Eq. \Ref{3.F6a} to obtain the low temperature expansion of the free energy in the corresponding case.

\section{Compilations of results}
In this section we represent the results for the low temperature expansions of the free energy for the specific models defined in Sect. 2. The relation between the low frequency expansion, Eq. \Ref{3.Fi2},   of a function $\Phi(\om)$, introduced in Eqs. \Ref{1.DF} or \Ref{3.F4}, and calculated in Sect. 4 for the specific models,  and the low temperature expansion of the free energy is given by Eq. \Ref{3.F6a}. Also, here we restored the dependence on the widths $a$ of the gap according to the rules given in Sect. 2, Eqs. \Ref{2.restore}-\Ref{2.restore2}. However, we did not restore the dependence on the constants $\hbar$, $c$ and $k_{\rm B}$ since that would overload the formulas and refer to the corresponding rules in Sect. 2, Eqs. \Ref{2.restore3} and \Ref{2.restore4}.
Also, in this section we do not consider the polarizations separately. All formulas are written for the \elm case, i.e., after adding the contributions from both polarizations. We follow the order of the models used before.

\subsection{Ideal conductor}
We start with the case of ideally conducting interfaces considered in Subsection 4.1. From Eq. \Ref{4.1.DF1} we get
\be\label{5.1} \Delta_T\F^{\rm id.cond.}=-\frac{\zeta_{\rm R}(3)}{2\pi}\,T^3+\frac{\pi^2}{45}\,a\,T^4+\dots\,,
\ee
where the dots denote exponentially decreasing contributions. As well known, the leading order is independent from $a$ and it does not contribute to the force.

\subsection{Fixed permittivity}
Here we consider dielectric medium with permittivity. For dielectrics behind both interfaces we get from Eq. \Ref{4.2.3.ee}
\bea\label{5.2}\Delta_T \F&=&
-\frac{(\ep-1)^2}{\ep+1}\,\frac{\zeta_{\rm R} (3)}{4 \pi } \, \,T^3
+ \left(\sqrt{\ep}-1\right)
    \left(\ep^2+\ep^{3/2}-2\right) \frac{ \pi ^2}{90} \, a \,T^4
\nn\\&&-    (\ep-1)^2 (\ep^2+1)\frac{3\zeta_{\rm R} (5)}{4\pi } \, a^2 \,T^5
\nn\\&&+ (\ep-1) (\ep-1) (24 \ep^3-36 \ep^2+29 \ep-6)\frac{ \pi ^4}{2835} \, a^3 \,T^6
\nn\\&&+ (\ep-1)^3 (\ep^2+1) \frac{15\zeta_{\rm R} (7)}{4\pi } \, a^4 \,T^7 +O\left(\,T^8\right).
\eea
In case of one interface ideally conducting we get from Eq. \Ref{4.2.3.ie}
\bea\label{5.3}\Delta_T \F&=&
-\frac{(\ep-1)^2}{\ep+1}\,\frac{\zeta_{\rm R} (3)}{8 \pi } \, \,T^3
+ \left(\ep^{3/2}(\ep-2)+1\right) \frac{ \pi ^2}{45} \, a \,T^4
\nn\\&&-   (\ep-1)^2 \ep^2\frac{3\zeta_{\rm R} (5)}{\pi } \, a^2 \,T^5
\nn\\&&+ (\ep-1)^2 \ep^{3/2}  (24 \ep^3-36 \ep^2+5 \ep+10)\frac{8 \pi ^4}{2835} \, a^3 \,T^6
\nn\\&&+ (\ep-1)^3 \ep^2 \frac{60\zeta_{\rm R} (7) \, a^4}{\pi }\,T^7  +O\left(\,T^8\right).
\eea
Like in the case of ideal conductors, the leading order does not contribute to the force. The first contribution to the force comes from the fourth order in $T$.
In  Eq. \Ref{5.2}, the coefficient in front of $T^4$ is, up to a factor,   $C_4$ in Eq. (12.97) in \cite{BKMM} and  the corresponding coefficient in  Eq. \Ref{5.3}  is, up to a factor,   $K_4$ in Eq. (15.17) in \cite{BKMM}. The contributions of orders $T^3$ and $T^4$ in Eq. \Ref{5.2} were obtained in \cite{geye05-72-085009,klim06-39-6495}. In \cite{geye05-72-085009} the coefficient $C_4$ is also calculated for dissimilar plates.

\subsection{The plasma model}
We start  with the case of media with different plasma frequencies, $\Oma$ and $\Omb$, behind the interfaces. Using Eq. \Ref{4.3.3.Fi} we get the expansion
\bea\label{5.9}\Delta_T   \F&=&
-\left(1+\frac{1}{a}\left(\frac{1}{\Oma}+\frac{1}{\Omb}\right)\right)\frac{\zeta_{\rm R}(3)}{2\pi}\,T^3
+\left(1+\frac{2}{a}\left(\frac{1}{\Oma}+\frac{1}{\Omb}\right)\right)\frac{\pi^2a}{45}\,T^4
\nn\\&&
+\left(1+\frac{1}{2}\left(\frac{\Omb}{\Oma}+\frac{\Oma}{\Omb}\right)
        -\frac{3}{a}\left(\frac{\Omb}{\Oma^2}+\frac{\Oma}{\Omb^2}\right)
\right.\nn\\&&\left.
        -\frac{3}{2a^2}\left(\frac{\Omb}{\Oma^3}+\frac{\Oma}{\Omb^3}+\frac{1}{\Oma^2}
        +\frac{1}{\Omb^2}\right)
\right) \frac{2\zeta_{\rm R}(5)}{\Oma\Omb\pi}\,T^5
\nn\\&&+\left(\frac{1}{\Oma^3} +\frac{1}{\Omb^3}\right)  \frac{8\pi}{105}\,T^6+O(T^7).
\eea
It is represented in a form that the behavior for large plasma frequencies can be seen easily. For both tending to infinity the ideal conductor case is recovered. The formula simplifies if the plasma frequencies behind the interfaces are equal, $\Oma=\Omb=\Om$,
\bea\label{5.10}\Delta_T   \F&=&
-\left(1+\frac{2}{a \Om} \right)\frac{\zeta_{\rm R}(3)}{2\pi}\,T^3
+\left(1+\frac{4}{a \Om} \right)\frac{\pi^2a}{45}\,T^4
\nn\\&&
+\left(1-\frac{3}{a \Om}-\frac{3}{a^2 \Om^2}\right) \frac{4\zeta_{\rm R}(5)}{\Om^2\pi}\,T^5
+   \frac{16\pi}{105\Om^3}\,T^6+O(T^7).
\eea
The case of one interface ideally conducting can be obtained from tending one of the plasma frequencies to infinity,
\bea\label{5.11}\Delta_T   \F&=&
-\left(1+\frac{1}{a \Om} \right)\frac{\zeta_{\rm R}(3)}{2\pi}\,T^3
+\left(1+\frac{2}{a \Om} \right)\frac{\pi^2a}{45}\,T^4
\nn\\&&
+\left(1-\frac{6}{a \Om}-\frac{3}{a^2 \Om^2}\right) \frac{ \zeta_{\rm R}(5)}{\Om^2\pi}\,T^5
+   \frac{8\pi}{105\Om^3}\,T^6+O(T^7),
\eea
and the index on the other frequency was dropped. The contributions of order $T^3$ to $T^5$ were obtained in \cite{geye01-16-3291}. These contributions, with account for \Ref{2.3.skin}, reproduce Eq. (14.14) in \cite{BKMM}, which was obtained as an expansion in powers of $\delta=1/\Om$. In general, the free energy for the plasma model has an asymptotic expansion in both, $T$ and $\delta$.

\subsection{The Drude model}
Here we consider the case of a fixed dissipation parameter $\ga$ (for vanishing $\ga$ see Sect. 4.4.2.) and restrict ourselves to the lowest order which is $T^2$, including $T^2\ln T$.   For medium behind both interfaces, adding the contributions from Eqs. \Ref{4.4.1.Fi2} and \Ref{4.4.1.Fi4}, we get
%
\be\label{5.8}\Delta_T   \F=\left[\frac{ \Om^2}{\ga}\left(2\ln2-1\right)
        -\frac{2\pi^2\ga}{3a^2\Om^2}\right]\frac{T^2}{48 }
    -\frac{\zeta_{\rm R}(5/2)\ \Om^3}{16\sqrt{2}\, \ga^{3/2}}\,aT^{5/2}
    +O(T^{3})
\ee
in agreement with \cite{hoye07-75-051127}. In case, one interface is ideally conducting, we have to add the contributions from Eqs. \Ref{4.4.1.Fit} and \Ref{4.4.1.Fi3} and we get
%
\bea\label{5.7}\Delta_T   \F&=&
    \left[\frac{ \Om^2}{96\ga}
        \left(\gamma_{\rm E} + \zeta_{\rm R}'(-1)-1+\frac12 \ln\frac{8a\Om^2}{\ga}\right)
            -\frac{\pi^2\ga}{3a^2\Om^2}\right]{T^2}
    +\frac{\Om^2}{96\ga}\ln(4\pi aT)\,{T^2}
\nn\\&&    -\frac{\zeta_{\rm R}(5/2)\Om^3}{8\sqrt{2}\pi^{3/2}\ga^{3/2}}aT^{5/2}
    +O(T^{3}),
\eea
which has the logarithmic contribution.

\subsection{Insulator described by oscillator model}

For an insulator described by the oscillator model, the expansions follow from Subsection 4.5. In case without dissipation, from Eq. \Ref{4.5.subst}, we get
\bea\label{5.4}\Delta_T \F&=&
-\frac{(\ep_0-1)^2}{\ep_0+1}\,\frac{\zeta_{\rm R} (3)}{4 \pi } \, \,T^3
+ \left(\sqrt{\ep_0}-1\right)
    \left(\ep_0^2+\ep_0^{3/2}-2\right) \frac{ \pi ^2}{90} \, a \,T^4
\nn\\&&-    (\ep_0-1)^2 \left(\ep_0^2+1+\frac{4}{(\ep_0+1)a^2\om_0^2}\right)
\frac{3\zeta_{\rm R} (5)}{4\pi } \, a^2 \,T^5
\nn\\\nn&& + (\ep_0-1)
 \left( (\ep_0+1) (24 \ep_0^3-36 \ep_0^2+29 \ep_0-6)
\right.\nn\\&&\left. -\frac{\sqrt{\ep_0}}{\ep_0+1}\,\frac{90\pi}{a^2\om_0^2}\right)
 \frac{ \pi ^4}{2835} \, a^3 \,T^6
+O\left(\,T^7\right),
\eea
and for one interface ideally conducting,
\bea\label{5.5}\Delta_T \F&=&
-\frac{(\ep_0-1)^2}{\ep_0+1}\,\frac{\zeta_{\rm R} (3)}{8 \pi } \, \,T^3
+ \left(\ep_0^{3/2}(\ep_0-2)+1\right) \frac{ \pi ^2}{45} \, a \,T^4
\nn\\&&-   (\ep_0-1)^2
\left(\ep_0^2+\frac{1}{(\ep_0+1)2a^2\om_0^2}  \right)
            \frac{3\zeta_{\rm R} (5)}{\pi } \, a^2 \,T^5
\nn\\\nn&&+ (\ep_0-1)
\left( \ep_0^{3/2}  (24 \ep_0^3-36 \ep_0^2+5 \ep_0+10)
\right.\nn\\&&\left.-\frac{\ep_0-1}{\ep_0+1}\,\frac{45\pi}{8a^2\om_0^2}  \right)
\frac{8 \pi ^4}{2835} \, a^3 \,T^6 +O\left(\,T^7\right).
\eea
These formulas differ from  \Ref{5.2} and \Ref{5.3} only by the substitution $\ep\to\ep_0$ and the contributions depending on $\om_0$, i.e., starting from the order $T^5$, which is quite trivial, but was not considered before.
The first two terms in  \Ref{5.4} and \Ref{5.5} were obtained in \cite{geye05-72-085009,klim06-39-6495}.

In case of dissipation we consider   the leading order,  which is $T^2$, and the order $T^4\ln T$ as  another example for logarithmic contributions in the low temperature expansion of the free energy. We did not calculate the order $T^4$ (without logarithm). For medium behind both interfaces we get from Table \ref{4.5.tab} adding the second and the fourth lines and using Eq. \Ref{3.zeta}
\be\label{5.6}\Delta_T \F=
-{\rm Li}_2\left(\frac{\ep_0-1}{\ep_0+1}\right) \,\frac{\ga_0}{(\ep_0^2-1)12a^2}\,T^2
-\frac{(\ep_0-1)(\ep_0+3)\ga_0}{(\ep_0+1)^2}\,\frac{\pi^4}{60}\ln(2aT) \,T^4+O(T^4),
\ee
and, for one interface ideally conducting, from adding the first and the third lines,
\be\label{5.5a}\Delta_T \F=
-{\rm Li}_2\left(\frac{\ep_0-1}{\ep_0+1}\right) \,\frac{\ga_0}{(\ep_0^2-1)24a^2}\,T^2
-\frac{(\ep_0-1)(\ep_0+3)\ga_0}{(\ep_0+1)^2}\,\frac{\pi^4}{30}\ln(2aT) \,T^4+O(T^4),
\ee
where we accounted for that $\ga_0$, Eq. \Ref{4.5.ga}, has dimension of one inverse frequency. The contributions from frequency dependence and dissipation are additive in the considered orders of the expansion. The logarithmic term contributes to the force, whereas contributions proportional to $\ga_0T^4$ (without logarithm) do not contribute due to a cancelation after restoration of the dependence on $a$.
Cross terms, depending on both, $\om_0$ and $\ga_0$, appear in higher orders, starting from $T^6$,  only. In the above expansion, in Eq. \Ref{5.6}, the order $T^2$ was obtained in \cite{klim08-41-164032}.

\subsection{The case of  dc conductivity}
We present here the results for a fixed conductivity $\sigma$, which is rather nonphysical, for completeness (for vanishing $\sigma$ see Sect. 4.6.2.). The low temperature expansion of the free energy follows, for medium behind both interfaces, from adding Eqs. \Ref{4.6.1.F2} and \Ref{4.6.1.F4},
%
\be\label{5.8d}\Delta_T   \F=
    \left[\frac{\pi  \sigma(\ln2-1)}{12}-\frac{\pi }{288a^2\sigma}\right]T^2
    -\frac{\zeta_{\rm R}(5/2)\,\sigma^{3/2}}{2\sqrt{2}}\,aT^{5/2}
    +O(T^{3}).
\ee
The contribution with $\sigma$ in the denominator results from the TM polarization. It was found in \cite{elli08-78-021117}.
For one interface ideally conducting we get from Eqs. \Ref{4.6.1.F1} and \Ref{4.6.1.F3},
%
\bea\label{5.7d}\Delta_T   \F&=&
    \left[\frac{\pi  \sigma}{24}\left(\gamma_{\rm E} + \zeta_{\rm R}'(-1)-1
        +\frac12\ln(32\pi a\sigma)\right)-\frac{\pi}{567a^2\sigma}\right]T^2
\nn\\&& +\frac{\pi\sigma}{24}\ln(4\pi aT)\,T^2
    -\frac{\zeta_{\rm R}(5/2)\,\sigma^{3/2}}{\sqrt{2}}\, aT^{5/2}
    +O(T^{3}),
\eea
which has a logarithmic contribution like Eq. \Ref{5.7} in the Drude model.
Both these expressions follow from \Ref{5.8} and \Ref{5.7} by the substitution $\frac{\Om^2}{\ga}\to4\pi\sigma$.

\subsection{Hydrodynamic model for graphene}
Here we have again two different plasma frequencies in the interfaces. From Eq.
\Ref{4.7.3.ED} we get the expansion
\bea\label{5.12}\Delta_T   \F&=&
-\left(1+\frac{1}{4a}\left(\frac{1}{\Oma}+\frac{1}{\Omb}\right)\right)\frac{\zeta_{\rm R}(3)}{2\pi}\,T^3
\nn\\&&
+\left[\frac{1+\frac{3}{2a}\left(\frac{1}{\Oma}+\frac{1}{\Omb}\right)
            +\frac{5}{8a^2}\left(\frac{1}{\Oma^2}+\frac{1}{\Omb^2}\right)
            +\frac{1}{a^2\Oma\Omb}}
            {1+\frac{1}{2a}\left(\frac{1}{\Oma}+\frac{1}{\Omb}\right)}
\right.\nn\\&&\left.
        +\frac{3}{2(2a)^{3/2}}\frac{\frac{1}{\Oma^2}+\frac{1}{\Omb^2}}
                                    {\sqrt{\frac{1}{\Oma}+\frac{1}{\Omb}}}
        \left(\frac{-\pi}{4\sqrt{2}}+\arctan\sqrt{\frac{2a\Oma\Omb}{\Oma+\Omb}}\right)
        \right] \frac{\pi^2 a}{45}\,T^4
\nn\\&&+\frac{246+88\left(\frac{\Omb}{\Oma}+\frac{\Oma}{\Omb}\right)
        +43\left(\frac{\Omb^2}{\Oma^2}+\frac{\Oma^2}{\Omb^2}\right)}
                    {(\Oma+\Omb)^2}  \frac{\zeta_{\rm R}(5)}{16\pi}\,T^5+O(T^6).
\eea
which is similar to Eq. \Ref{5.9} in the plasma model. The special case of equal plasma frequencies is
\bea\label{5.13}\Delta_T   \F&=&
-\left(1+\frac{1}{2a\Om}\right)\frac{\zeta_{\rm R}(3)}{2\pi}\,T^3
\nn\\&&
+\left[\frac{1+\frac{3}{a\Om}+\frac{9}{4a^2\Om^2}}{1+\frac{1}{a\Om}}
        +\frac{3}{4(a\Om)^{3/2}}
        \left(\frac{-\pi}{4\sqrt{2}}+\arctan\sqrt{a\Om}\right)
        \right] \frac{\pi^2 a}{45}\,T^4
\nn\\&&+\frac{127\zeta_{\rm R}(5)}{16\pi\Om^2}\,T^5
-\left[ \frac{ 1+\frac{39}{64a\Om}-\frac{101}{24a^2\Om^2}-\frac{407}{64 a^3\Om^3}
            -\frac{5}{2a^4\Om^4}}{\left(1+\frac{1}{a\Om}\right)^3}
\right.\\\nn&&\left.        +\frac{25}{64\sqrt{a\Om}}
            \left(\frac{\pi}{8\sqrt{2}(1+\frac{1}{a\Om})^3}
                    -\arctan\sqrt{a\Om}\right)  \right]  \frac{4\pi^4a}{315 \Om^2}\,T^6
+O(T^7).
\eea
Again we see that for $\Om\to\infty$ the ideal conductor case is recovered. Finally we note the case if one interface is ideally conducting,
\bea\label{5.14}\Delta_T   \F&=&
-\left(1+\frac{1}{4a\Om}\right)\frac{\zeta_{\rm R}(3)}{2\pi}\,T^3
\nn\\&&
+\left[\frac{1+\frac{3}{2a\Om}+\frac{5}{8a^2\Om^2}}{1+\frac{1}{2a\Om}}
        +\frac{3}{2(2a\Om)^{3/2}}
        \left(\frac{-\pi}{4\sqrt{2}}+\arctan\sqrt{a\Om}\right)
        \right] \frac{\pi^2 a}{45}\,T^4
\nn\\&&+\frac{43\zeta_{\rm R}(5)}{16\pi\Om^2}\,T^5
-\left[ \frac{1+\frac{91}{96 a\Om}-\frac{71}{24a^2\Om^2}-\frac{231}{128 a^3\Om^3}
            -\frac{1}{2a^4\Om^4}}{\left(1+\frac{1}{2a\Om}\right)^3}
\right.\\\nn&&\left.        +\frac{25}{16\sqrt{2a\Om}}
            \left(\frac{\pi}{8\sqrt{2}}
                    -\arctan\sqrt{a\Om}\right)  \right]  \frac{2\pi^4a}{315 \Om^2}\,T^6
+O(T^7).
\eea
It is interesting to mention that the order $T^5$ does not depend on the width $a$.

\section{Conclusions}
In the forgoing sections we considered the asymptotic expansion of the free energy as given by the Lifshitz formula for low temperature. We used the Abel-Plana formula and exploited, in a unified treatment,  the analytic properties of the reflection coefficients. Of course, we reproduced the known results, in a number of cases we were able to go beyond.

For instance, in the plasma model, it is possible to get arbitrarily many terms of the expansion in a systematic way. In other models, more elaborate methods are necessary to calculate higher orders in the asymptotic expansions of some integrals. The basic tool, used in this paper, is Eq. \Ref{1.DF} together with Eqs. \Ref{1.Fa} and \Ref{1.fi}. Together with the division \Ref{3.reg} into regions, this allows for some more insight into the structure of the free energy at low temperature. So, there are  from region \a\, contributions with,  besides real frequency $\om$,  real wave numbers $q$ and $k_3$, which correspond to scattering states in the spectrum. From region \b\, there are contributions with real frequency and imaginary wave number, $q=i\eta$ (inside the gap),  which correspond to surface plasmons. In this way, the low temperature expansion is determined by the low lying excitations in the spectrum. In this sense, the expansions for the specific models, considered in this paper, can be viewed as illustration to the general formulas  in Chapter 5.2 in \cite{BKMM}.

It must be mentioned that this picture becomes significantly more complicated if including dissipation, which brings its own contribution to the imaginary part and contributions from all frequencies, including high ones, play  a role. An example is the Drude model in Sect. 4.11 with fixed dissipation parameter $\ga$, where in case 1, in order to obtain the expansions \Ref{4.4.1.fi4}, Eq. \Ref{A.1} from Appendix A was used. To that formula  arbitrary high wave numbers contribute.

In this paper, in the cases with dissipation, i.e., the Drude model and  in the case with dc conductivity, we re-derived the contributions violating the third law of thermodynamics. These models are at once non perturbative in the respective dissipation parameter in the sense that, for vanishing dissipation parameter, a contribution is left in addition to what one calculates starting from no dissipation. These contributions are  derived in the original way starting from the Matsubara representation, and in an alternative way starting from representation \Ref{1.DF} in terms of real frequencies. Since we do not enter here the discussion on the applicability of these models together with the Lifshitz formula, we restrict ourselves to the following remark.
As known, and as can be seen from Eq. \Ref{5.2}, the limit $T\to0$ and $\ep\to\infty$ (where  in the reflection coefficient the ideal conductor case is recovered), do not commute. A similar non-commutativity we observe  in the case with  the dissipation parameters, $\ga$ or $\sigma$, tending to zero. So, for instance, in the Drude model (TE case), Sect. 4.4.1, Eqs. \Ref{4.4.1.Fi} and \Ref{4.4.1.Fi2}, and  in the dc model (TM case) in Sect. 4.6, Eqs. \Ref{4.6.1.F3} and \Ref{4.6.1.F4}, have the dissipation parameter in the denominator and the limit of tending the dissipation parameter to zero does not commute with the limit $T\to0$.

We included also the hydrodynamic model for graphene, Sect. 4.7. This model is very much like the plasma model, which is quite natural since both can be derived from a charged fluid, one time confined to a half space, the other time to a plane. So, in the low temperature expansion, we see similar structures, especially concerning the role of the surface plasmons. As for the applicability of these results to graphene we mention that its low frequency properties are better described by the Dirac model which is not included in this paper.

Finally we mention that the considered models are taken as such. We do not bother where these models come from. What we did is to consider the low temperature expansion of the Lifshitz formula with these models inserted. In fact, these models appear from the dynamics of the electrons in the bodies, whose Casimir interaction is described by the Lifshitz formula. In general, these models have their own temperature dependence. In this sense, we considered the photons at temperature $T$, whereas the electrons are taken at zero temperature. It is, in fact, only the ideal conductor model, where this is justified by discussing  that the plates are ideal at any temperature.  For other models one could speculate, that a gap in the electronic excitations (like the plasma frequency in the plasma model) would exponentially suppress the corresponding temperature dependence. For the Drude model and for a dielectric with fixed dc conductivity, we considered the case where the dissipation parameter depends on temperature. This is, so to say, the simplest way to account for the temperature. It results in the known, formal at least, violation of the third law. A more complete account for the temperature dependence is, of course, interesting, but beyond the scope of this paper.

\section*{Acknowledgements}
The authors is highly indebted to valuable discussions with G. Klimchitskaya and V. Mostepanenko.

\appendix
\section{Asymptotic expansion of an integral}
In this Appendix we calculate the asymptotic expansion of an integral appearing in Sect.s 4, Eqs. \Ref{4.4.1.fi1} and \Ref{4.4.2.F3}, and in Sect. 6, Eq. \Ref{4.6.1.fi} for $k=1$. It can be written in the form
\be\label{A.1}I(\Omega)=\int_0^\infty d\eta\,\eta\,
            \ln\left(1-r_0r_1(\Omega)e^{-\eta}\right),
\ee
where $r_0$ is a number and
\be\label{A.r1Om}r_1(\Omega)=\frac{\eta-\sqrt{\Omega^2+\eta^2}}{\eta+\sqrt{\Omega^2+\eta^2}}
\ee
was introduced in Eq. \Ref{4.4.2.Om}.

We are interested in an expansion for   $\Omega \to0$ up to some power $\Omega^{n_0}$,
\be\label{A.2}I(\Omega)=\sum_{i=0}^{n_0}
    a_i(r_0)\left(\frac{\Omega}{2}\right)^i+O(\Omega^{n_0+1}).
\ee
A direct expansion of the integrand is not possible since it would be in powers of $\Omega/\eta$ resulting in a singular integration at $\eta\to0$. The same way an expansion after the substitution $\eta\to\eta\om$ is not possible since it would result in singularities for $\eta\to\infty$.

We proceed as follows. First we make a substitution,
\be\label{A.subst}\eta=\frac{\Omega}{2}(t-t^{-1}),
\ee
in \Ref{A.1},
\be\label{A.3}I(\Omega)=\frac{\Omega^2}{4}\int_1^\infty dt\,t\,(1-t^{-4})\,\ln\left(1+\frac{r_0}{t^2}\,e^{-\frac{\Omega}{2}(t-1/t)}\right).
\ee
Next we assume $|r_0|<1$ and expand the logarithm,
\be\label{A.4}I(\Omega)=\sum_{s=1}^\infty \frac{-r_0^s}{s}A_s(\Omega)
\ee
with
\be\label{A.A1}A_s(\Omega)=\left(\frac{\Omega}{2}\right)^2 \int_1^\infty
dt\,t^{1-2s}\,(1-t^{-4}) \,e^{-\frac{s\Omega}{2}(t-1/t)}.
\ee
In this representation, for small $\Omega$, the main contribution comes from large $t$. Thus we can expand the exponential
\be\label{A.exp}\exp\left({\frac{s\Omega}{2t}}\right)=\sum_{k=0}^{k_0}
\frac{1}{k!}\left(\frac{s\Omega}{2t}\right)^k+\dots\,,
\ee
which gives $k$ additional powers of $\Omega$ in excess of the factor $\Omega^2$ in front of the integral and it is sufficient to take $k_0=n_0-2$. We insert this expansion into \Ref{A.A1},
\be\label{A.A2}A_s(\Omega)=\left(\frac{\Omega}{2}\right)^2 \int_1^\infty
dt\,t^{1-2s}\,(1-t^{-4}) \,e^{-\frac{s\Omega}{2}t}\sum_{k=0}^{k_0}
\frac{1}{k!}\left(\frac{s\Omega}{2t}\right)^k+O(\Omega^{n_0+1}).
\ee
We split this integral into two by adding and subtracting the integral from $t=0$ till $t=1$,
\be\label{A.A3}A_s(\Omega)=P_s(\Omega)-Q_s(\Omega)
\ee
with
\be\label{A.P1}P_s(\Omega)=\left(\frac{\Omega}{2}\right)^2 \int_0^\infty
dt\,t^{1-2s}\,(1-t^{-4}) \,e^{-\frac{s\Omega}{2}t}
\sum_{k=0}^{k_0}
\frac{1}{k!}\left(\frac{s\Omega}{2t}\right)^k+O(\Omega^{n_0+1})
\ee
and
\be\label{A.Q1}Q_s(\Omega)=\left(\frac{\Omega}{2}\right)^2 \int_0^1
dt\,t^{1-2s}\,(1-t^{-4}) \,e^{-\frac{s\Omega}{2}t}
\sum_{k=0}^{k_0}
\frac{1}{k!}\left(\frac{s\Omega}{2t}\right)^k+O(\Omega^{n_0+1}).
\ee
Making the analytic continuation to sufficiently small $s$, the integral in \Ref{A.P1}  can be done,
\be\label{A.P2}P_s(\Omega)=\sum_{k=0}^{k_0}
\frac{s^{2(s+k-1)}}{k!}\left(\frac{\Omega}{2t}\right)^{2(s+k)}
\left(\Gamma(2-2s-k)-\left(\frac{s\Omega}{2}\right)^4\Gamma(-2-2s-k)\right).
\ee
The split in Eq. \Ref{A.A3} is possible for $s<1-\frac{k_0}{2}$. Eq. \Ref{A.P2} provides the conti\-nuation to the whole $s$-plane. The poles appearing in $s=\frac{2-k-N}{2}$, $N=0,1,\dots$, will be compensated by corresponding poles in $Q_s(\Omega)$, leaving logarithmic contributions behind. These poles appear for $s=0,1,\dots,s_0$ with $s_0=[\frac{n_0}{2}]$. All relevant powers of $\Omega$ come in $P_s(\Omega)$ from $s\le s_0$.

In $Q_s(\Omega)$, \Ref{A.Q1}, because of $t\le 1$, we can expand the remaining exponential,
\be\label{A.Q2}Q_s(\Omega)=\left(\frac{\Omega}{2}\right)^2 \int_0^1
dt\,t^{1-2s}\,(1-t^{-4}) \,
\sum_{k=0}^{k_0}
\frac{1}{k!}\left(\frac{s\Omega}{2}(t-t^{-1})\right)^k+O(\Omega^{n_0+1}),
\ee
and the integration becomes simple allowing for an easy machinized treatment. With these representations for $P_s(\Omega)$ and $Q_s(\Omega)$, we return to Eq. \Ref{A.4}.

We split the sum,
\be\label{A.AB}I(\Omega)=A+B+O(\Omega^{n_0+1}),
\ee
into
\be\label{A.A}A=\sum_{s=1}^{s_0}\frac{-(r_0)^s}{s}\left(P_s(\Omega)-Q_s(\Omega)\right)
\ee
and
\be\label{A.B}B=\sum_{s_0+1}^{\infty}\frac{(r_0)^s}{s}\,Q_s(\Omega).
\ee
The sum in $A$ is a finite one  and that in $B$ is over a rational function of $s$. For any given $n_0$, these can be easily calculated machinized. In doing so one can generate expression for the coefficients $a_i(r_0)$ in \Ref{A.2}. The first two coefficients are zero, $a_0(r_0)=0$ and $a_1(r_0)=0$. The next two are
\bea\label{A.a}
a_2(r_0)&=&\frac{1-r_0}{2}+\frac{(1-r_0)^2}{2r_0}\ln(1-r_0)+r_0\ln\frac{\Omega}{2},
\nn\\
a_3(r_0)&=&\frac{2}{3}-\frac{1+r_0}{r_0}
    +\frac{(1+r_0)(1-r_0)^2}{r_0^{3/2}}{\rm arctanh}\sqrt{r_0}+\gamma_{\rm E}.
\eea
Higher orders can be easily generated.
For special cases for $r_0$ the expressions simplify. For $r_0=-1$ we get
\be\label{A.r-1}\begin{array}{cclccr}
                 a_2(-1)&=&  -\log (2 \Omega )-\gamma_{\rm E} +1,  & a_3(-1)&=& \frac{8}{3}, \\ [8pt]
                 a_4(-1)&=&2 \log (\Omega )+\log (2)+2 \gamma_{\rm E} -3, &  a_5(-1)&=&-\frac{32}{9}, \\[8pt]
   a_6(-1)&=& -\frac{8}{3} \log \left(\frac{\Omega }{2}\right)-\frac{9 \log (3)}{8}-4 \log (2)-\frac{8 \gamma_{\rm E}   }{3}+\frac{91}{18} ,&
a_7(-1)&=&\frac{128}{2},
               \end{array}
\ee
for $r_0=0$ we get
\be\label{A.r0}\begin{array}{cclccr}
a_2(0)&=& \log \left(\frac{\Omega }{2}\right)+\gamma_{\rm E} +\frac{1}{4},
& a_3(0)&=& -\frac{32}{15}, \\ [8pt]
a_4(0)&=& -\log (\Omega )+\log
   (2)-\gamma_{\rm E} +\frac{13}{12}, &
a_5(0)&=&\frac{256}{315}, \\[8pt]
a_6(0)&=& \frac{1}{576} \left(120 \log \left(\frac{\Omega
   }{2}\right)+120 \gamma_{\rm E} -191\right),&
a_7(0)&=&-\frac{512}{4725},
               \end{array}
\ee
and for $r_0=1$ it is
\be\label{A.r1}\begin{array}{cclccr}
a_2(1)&=&  \log \left(\frac{\Omega }{2}\right)+\gamma_{\rm E}  ,
& a_3(1)&=&  -\frac{4}{3}, \\ [8pt]
a_4(1)&=&\log (2) , &
a_5(1)&=& -\frac{4}{9}, \\[8pt]
a_6(1)&=&  -\frac{4   \log (2)}{3}+\frac{9 \log (3)}{8},&
a_7(1)&=&-\frac{52}{225},
               \end{array}
\ee
where $\gamma_{\rm E}$ is Euler's constant.

\section{The $\om$-integration in Eqs. \Ref{4.4.2.A2} and \Ref{4.6.2.A2}}
We consider the integral
\be\label{B.1}I=\int_0^\infty\frac{d\om}{\om}\, i\left(\phi(i \om)-\phi(-i \om)\right)
\ee
as it appears in Eqs. \Ref{4.4.2.A2} and \Ref{4.6.2.A2}. Since the functions $\phi(\pm i\om)$ result from Eqs. \Ref{4.4.2.fi0} and \Ref{4.6.1.fi}, we can be sure that these do not have singularities in the upper half plane, $\Re \om>0$.  Also we assume, that these functions take finite values in $\om=0$ and for $\om\to\infty$.

The integration in \Ref{B.1} is converging for $\om\to0$ and for $\om\to\infty$  due to the cancelations which take place in the parentheses. We  would like to consider both contributions separately. For the corresponding integrals to converge we introduce regularizations on both ends and represent
\be\label{B.2}I=i\lim_{r\to0\atop R\to\infty}\left[
    \int_r^R\frac{d\om}{\om}\, \phi(i \om)-
    \int_r^R\frac{d\om}{\om}\,  \phi(-i \om)  \right].
\ee
In the first integral we change the integration variable, $\om\to-\om$. Further we add and subtract integrals along pathes $\gamma_r$ and $\gamma_R$. Both are half circles in the upper half of the imaginary plane with radii $r$ and $R$ such that
\be\label{B.ga}\gamma=[-R,-r]\cup\gamma_r\cup[r,R]\cup\gamma_R
\ee
is a closed path. Accordingly, we rewrite the integrals in the form
\be\label{B.3}I=i\lim_{r\to0\atop R\to\infty}\left[
 \int_{\gamma}\frac{d\om}{\om}\, \phi(-i \om)
 -\int_{\gamma_r}\frac{d\om}{\om}\, \phi(-i \om)
 -\int_{\gamma_R}\frac{d\om}{\om}\, \phi(-i \om)        \right].
\ee
Now, for the functions we are considering, the integral over the closed path is zero. From the other two integrals, in the limit, we get contributions from the half poles,
\be\label{B.4}I=\pi\left(\phi(0)-\phi(\infty)\right),
\ee
which is the desired formula.

\end{document}